\pdfoutput=1

\documentclass[11pt]{article}

\usepackage[final]{acl}

\usepackage{times}
\usepackage{latexsym}

\usepackage[T1]{fontenc}

\usepackage[utf8]{inputenc}

\usepackage{microtype}

\usepackage{inconsolata}

\usepackage{graphicx}

\usepackage{amstext}
\usepackage{amsmath}
\usepackage{hyperref}
\usepackage{cleveref}
\crefname{section}{§}{§§}
\Crefname{section}{§}{§§}
\usepackage{threeparttable}
\usepackage{algorithm2e}
\usepackage{makecell}
\usepackage{booktabs}
\usepackage{stfloats}
\usepackage{multirow}
\usepackage{bbding}
\usepackage{rotating}
\usepackage{tabularx}
\usepackage{colortbl}
\usepackage{wrapfig}
\usepackage{setspace}
\usepackage{algpseudocode}
\usepackage{graphicx}
\usepackage{pifont}
\usepackage{enumitem}

\newcommand{\Logo}{\raisebox{-0.35\height}{\includegraphics[width=2em]{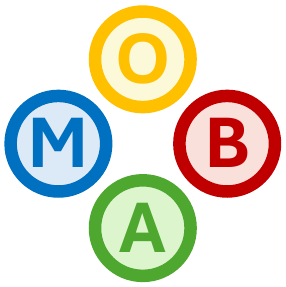}}}
\newcommand{\agent}{\textsc{MobA}}
\newcommand{\bench}{\textsc{MobBench}}

\usepackage{ifthen}
\usepackage{ulem}
\usepackage{CJKutf8}

\newif\ifshowcomments
\showcommentsfalse

\newcommand{\COMMENTbyL}[2][]{
\begin{CJK}{UTF8}{gbsn}\ifshowcomments
  \ifthenelse{\equal{#1}{done}} 
    {\textcolor{blue}{(\sout{<Solved> Lu}: #2)}}
    {\ifthenelse{\equal{#1}{key}}
        {\textcolor{blue}{\textbf{(Lu: #2)}}} 
        {\textcolor{blue}{(Lu: #2)}}}
        \fi\end{CJK}}

\newcommand{\COMMENTbyK}[2][]{
\begin{CJK}{UTF8}{gbsn}\ifshowcomments
  \ifthenelse{\equal{#1}{done}} 
    {\textcolor{red}{(\sout{<Solved> Kai}: #2)}}
    {\ifthenelse{\equal{#1}{key}}
        {\textcolor{red}{\textbf{(Kai: #2)}}} 
        {\textcolor{red}{(kai: #2)}}}
        \fi\end{CJK}}

\newcommand{\ZC}[2][]{
\begin{CJK}{UTF8}{gbsn}\ifshowcomments
  \ifthenelse{\equal{#1}{done}} 
    {\textcolor[rgb]{0.25, 0.75, 0.25}{(\sout{<Solved> Zichen}: #2)}}
    {\ifthenelse{\equal{#1}{key}}
        {\textcolor[rgb]{0.25, 0.75, 0.25}{\textbf{(Zichen: #2)}}} 
        {\textcolor[rgb]{0.25, 0.75, 0.25}{(Zichen: #2)}}}
        \fi\end{CJK}}

\newcommand{\COMMENTbyLT}[2][]{
  \begin{CJK}{UTF8}{gbsn}\ifshowcomments
  \ifthenelse{\equal{#1}{done}} 
    {\textcolor[rgb]{0.25, 0.25, 0.75}{(\sout{<Solved> Liangtai}: #2)}}
    {\ifthenelse{\equal{#1}{key}}
        {\textcolor[rgb]{0.25, 0.25, 0.75}{\textbf{(Liangtai: #2)}}} 
        {\textcolor[rgb]{0.25, 0.25, 0.75}{(Liangtai: #2)}}}
  \fi\end{CJK}}

\newcommand{\COMMENTbyTH}[2][]{
  \begin{CJK}{UTF8}{gbsn}\ifshowcomments
  \ifthenelse{\equal{#1}{done}} 
    {\textcolor[rgb]{0.5, 0.5, 0.25}{(\sout{<Solved> Tanghao}: #2)}}
    {\ifthenelse{\equal{#1}{key}}
        {\textcolor[rgb]{0.5, 0.5, 0.25}{\textbf{(Tanghao: #2)}}} 
        {\textcolor[rgb]{0.5, 0.5, 0.25}{(Tanghao: #2)}}}
  \fi\end{CJK}}

\newcommand{\COMMENTbyST}[2][]{
  \begin{CJK}{UTF8}{gbsn}\ifshowcomments
  \ifthenelse{\equal{#1}{done}} 
    {\textcolor[rgb]{0.2, 0.65, 0.6}{(\sout{<Solved> Situo}: #2)}}
    {\ifthenelse{\equal{#1}{key}}
        {\textcolor[rgb]{0.2, 0.65, 0.6}{\textbf{(Situo: #2)}}} 
        {\textcolor[rgb]{0.2, 0.65, 0.6}{(Situo: #2)}}}
  \fi\end{CJK}}  

\newcommand{\CDY}[2][]{
  \begin{CJK}{UTF8}{gbsn}\ifshowcomments
  \ifthenelse{\equal{#1}{done}} 
    {\textcolor[rgb]{0.2, 0.65, 0.6}{(\sout{<Solved> Danyang}: #2)}}
    {\ifthenelse{\equal{#1}{key}}
        {\textcolor[rgb]{0.2, 0.65, 0.6}{\textbf{(Danyang: #2)}}} 
        {\textcolor[rgb]{0.2, 0.65, 0.6}{(Danyang: #2)}}}
  \fi\end{CJK}}

\title{\Logo~\agent{}: Multifaceted Memory-Enhanced Adaptive Planning for Efficient Mobile Task Automation}

\author{
        Zichen Zhu, Hao Tang, Yansi Li, Dingye Liu, Hongshen Xu, 
        \\\textbf{Kunyao Lan, Danyang Zhang, Yixuan Jiang, Hao Zhou, Chenrun Wang},
        \\\textbf{Situo Zhang, Liangtai Sun, Yixiao Wang, Yuheng Sun, Lu Chen\footnotemark[1], Kai Yu\footnotemark[1]}\\
          X-LANCE Lab, Department of Computer Science and Engineering \\ MoE Key Lab of Artificial Intelligence, SJTU AI Institute \\ Shanghai Jiao Tong University, Shanghai, China \\
     \texttt{\{JamesZhutheThird,chenlusz,kai.yu\}@sjtu.edu.cn}
}

\begin{document}
\maketitle

\renewcommand{\thefootnote}{\fnsymbol{footnote}}
\footnotetext[1]{Corresponding authors are Lu Chen and Kai Yu.}
\renewcommand{\thefootnote}{\arabic{footnote}}

\begin{abstract}

Existing Multimodal Large Language Model (MLLM)-based agents face significant challenges in handling complex GUI (Graphical User Interface) interactions on devices. These challenges arise from the dynamic and structured nature of GUI environments, which integrate text, images, and spatial relationships, as well as the variability in action spaces across different pages and tasks. To address these limitations, we propose \agent{}, a novel MLLM-based mobile assistant system. \agent{} introduces an adaptive planning module that incorporates a reflection mechanism for error recovery and dynamically adjusts plans to align with the real environment contexts and action module's execution capacity. Additionally, a multifaceted memory module provides comprehensive memory support to enhance adaptability and efficiency. We also present \bench{}, a dataset designed for complex mobile interactions. Experimental results on \bench{} and AndroidArena demonstrate \agent{}'s ability to handle dynamic GUI environments and perform complex mobile tasks. 

\end{abstract}

\section{Introduction}

Multimodal large language models (MLLMs) have seen significant advancements in recent years, supported by vast multimodal datasets. These models~\cite{hu2023viscpm,liu2024improved,ye2024mplugowlmodularizationempowerslarge,ye2023mplugowl2revolutionizingmultimodallarge,chen2024internvlscalingvisionfoundation,sun2024generativemultimodalmodelsincontext,NEURIPS2023_6dcf277e,NEURIPS2023_9a6a435e,chen2023minigptv2largelanguagemodel,zhu2024minigpt,yao2024minicpmvgpt4vlevelmllm,gpt4v,team2023gemini} excel in tasks such as Chain-of-Thought (CoT) reasoning~\cite{NEURIPS2022_9d560961}, In-Context Learning (ICL)~\cite{NEURIPS2020_1457c0d6}, and various applications~\cite{wang2024exploringreasoningabilitiesmultimodal,wang2023geminireasoningunveilingcommonsense,chen2024mllmstrongrerankeradvancing,liu2024hiprompttuningfreehigherresolutiongeneration,pan2024kosmosg,ge2024seeddataedittechnicalreporthybrid,wu2024visionllmv2endtoendgeneralist,lee2024llavadocentinstructiontuningmultimodal,qian-etal-2024-chatdev,qian-etal-2024-experiential}. Their capabilities have also enabled new MLLM-based agents for real-world tasks~\cite{li2017sugilite,li2019pumice,sun2022meta,zhu2023cam,zhan2023autoui,yang2023appagent,zhang2024ufo,nong2024mobileflowmultimodalllmmobile,ma-etal-2024-coco,wang2024mobile,wang2024mobile2}.

However, MLLMs face significant challenges when addressing complex GUI interactions and facing diverse user demands in real-world scenarios, particularly on devices such as smartphones~\cite{zhang2024mobileenvbuildingqualifiedevaluation} and computers~\cite{cao2024spider2vfarmultimodalagents,xie2024osworldbenchmarkingmultimodalagents}. 
On the one hand, GUI environments are highly diverse and pose different action spaces across different apps and pages. For instance, the number and position of clickable icons can vary greatly across pages; some pages require text input, while others involve scrollable elements. Such variability makes proactive task planning hardly adapt to the real environment contexts and thus become infeasible to complete. 
On the other hand, the action executor can also lack capabilities enough to achieve it, even given a feasible task plan. 
In all these cases, agents with trivial or static planning~\cite{BoyuanZheng2024ICML_SeeAct,zhang2024ufo,nong2024mobileflowmultimodalllmmobile,ma-etal-2024-coco,xing2024understanding} will fail to align with the environment contexts and action executor's capacity and thus can fail the whole task easily caused by failure of a single sub-task.
Furthermore, existing MLLM-based GUI agents~\cite{DanyangZhang2023NeurIPS_Rememberer,yang2023appagent,wang2024mobile,wang2024mobile2} often lack a powerful and comprehensive memory to face the need for dynamic planning at various levels and diverse user demands.
These problems hinder the design of a practical mobile assistant.

    \begin{figure*}[!t]
      \centering
      \includegraphics[width=1.0\textwidth,trim=0 0 0 0,clip]{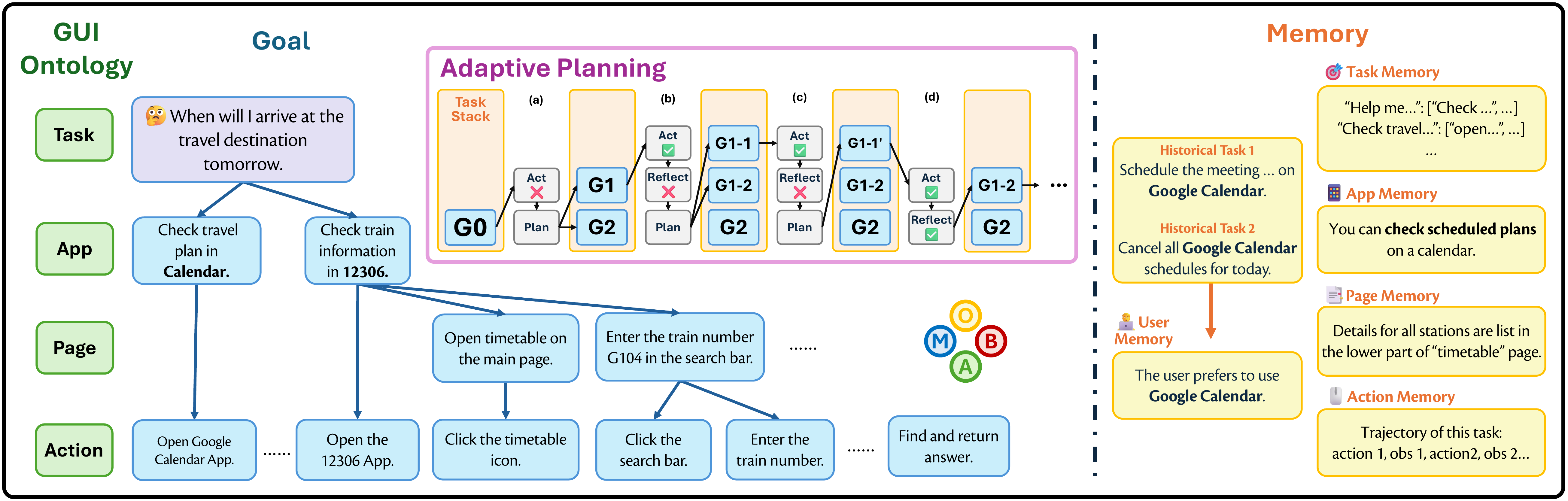}
      \caption{\textbf{The Illustration of adaptive planning and multifaceted memory structure.} There are 4 cases in adaptive planning: (a) Plan reflection failure, the goal needs to be decomposed. (b) In execution reflection failure, the goal needs to be decomposed. (c) Execution reflection failure, the goal needs to be refined. (d) Goal complete.}
      \label{fig:new_pipeline}
        \vspace{-0.3cm}
    \end{figure*}

To address these challenges, we propose \agent{}, a novel MLLM-based mobile assistant system with an adaptive planning module that dynamically adjusts task plans according to the execution results. As proactive planning often fails to accurately determine the actions required in a specific application or page or to align with the action executor's capacity, \agent{} leverages reflection mechanism to recover task execution from failed sub-plans by reassessing goals or breaking tasks into more fine-grained sub-goals. To better support adaptive planning with various sub-goal granularity, a multifaceted memory module providing hierarchical memory support is proposed. We also introduce \bench{}, a diverse dataset for complex mobile interactions, and demonstrate \agent{}'s effectiveness on \bench{} and AndroidArena~\cite{xing2024understanding}, showing its capability to handle dynamic GUI environments.

Our contributions are threefold:
\begin{itemize}[itemsep=4pt,topsep=2pt,parsep=0pt,nolistsep]
\item We propose an \textbf{adaptive planning} module that incorporates a reflection mechanism for error recovery and dynamically adjusts plans based on the current GUI environment and action executor's capacity.
\item We develop a \textbf{multifaceted memory} module that provides hierarchical memory support to enhance task adaptability and efficiency.
\item We introduce \textbf{\bench{}}, a diverse dataset for complex mobile interactions, and validate the effectiveness of our approach through extensive experiments on two datasets.
\end{itemize}

\section{The \agent{} System}

\begin{figure}[!t]
\centering
  \includegraphics[width=1.0\linewidth,trim=0 0 0 0,clip]{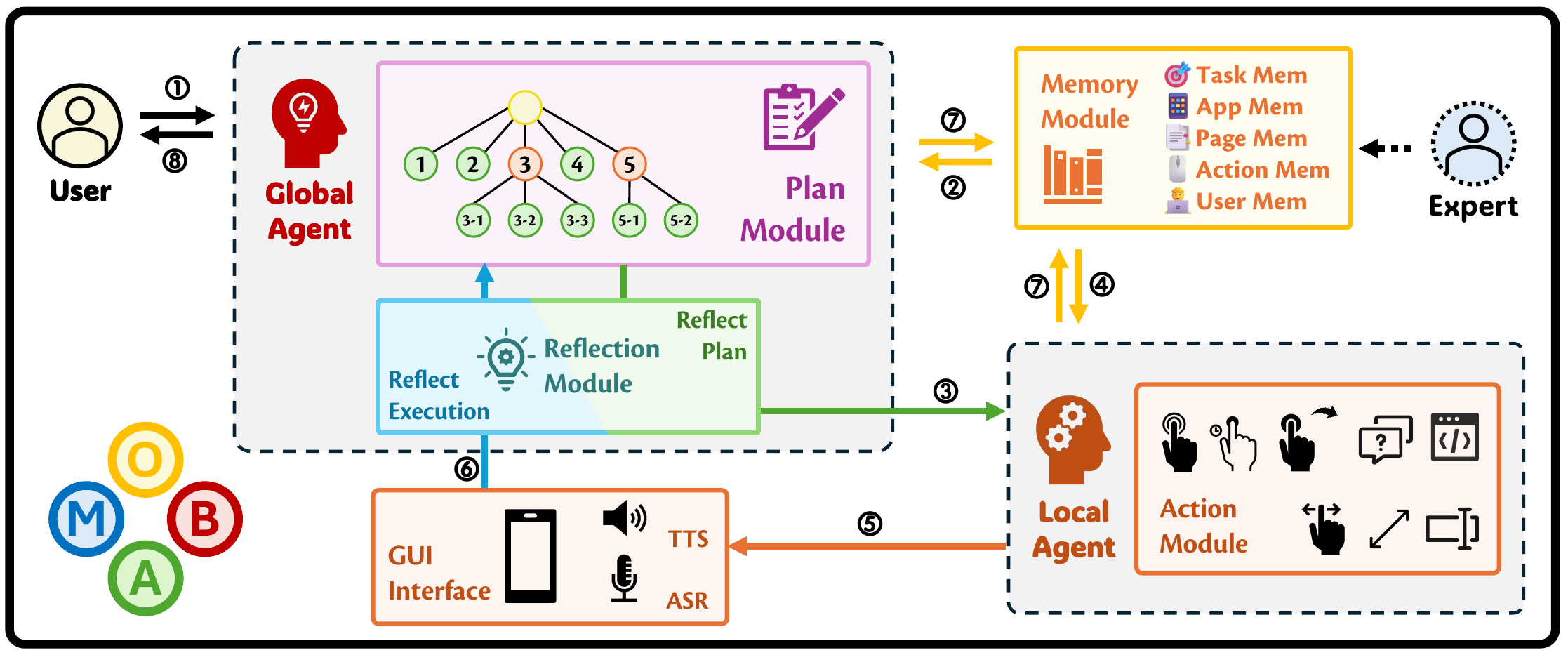}
        \caption{\textbf{System Overview of \agent.}  
        }
        \label{fig:overview}
        \vspace{-0.5cm}
    \end{figure}

The system overview of \agent{} is shown in \Cref{fig:overview}. \agent{} comprises a \textbf{Global Agent} (GA) and a \textbf{Local Agent} (LA). The Global Agent consists of a \uline{Plan Module} and a \uline{Reflection Module}. The Plan Module interprets the user's command (\ding{192}) and resolves the task into several easier and clearer sub-tasks adaptively with the help of experiences in the multifaceted \uline{Memory Module} (\ding{193}), while the Reflection Module will review if the decomposition is feasible and sub-goals are achievable. Then, under the direction of a specific sub-goal (\ding{194}), the Local Agent will leverage the experiences in the Memory Module (\ding{195}), predict the concrete actions, and directly control the device (\ding{196}). After LA's execution, the Reflection Module will reflect if the current sub-task has been completed (\ding{197}) and the Plan Module can revise the plan accordingly. The Memory Module can also be updated after the invocation of the Plan Module and Local Agent (\ding{198}) to improve \agent{}'s performance through execution. To ameliorate the performance at the early stage of the memory, it can also be initialized with a warm-up of some basic expert experiences. Finally, \agent{} can generate a response to the user regarding the result of task execution (\ding{199}). The remaining parts of this section will elaborate on the proposed \emph{adaptive} plan module and \emph{multifaceted} memory module.

\subsection{Task Completion with Adaptive Planning}

Facing the problem that static fixed-level task planning is deficient in aligning with real environment contexts and the Action Module's capacity, we propose adaptive planning to react to concrete execution results of the Action Module and adjust the granularity of task decomposition adaptively. The proposed planning workflow is demonstrated in Algorithm~\ref{alg:outline}. Given an established sub-goal, the reflection module is first adopted to review the sub-task feasibility. Then the Action Module will attempt to complete the reviewed sub-goal. The execution result will be inspected again by the Reflection Module. Once failure is detected, the Plan Module is invoked to revise the task plan to adapt to the current environment context or to further break the sub-goal down to match the Action Module's execution capacity. By repeating this procedure, \agent{} can generate a multi-granularity task plan that well aligns with the environment contexts and the Action Module's capacity iteratively and dynamically.

\begin{algorithm}[h]
    \small
    \rule{0.9\linewidth}{1pt}
    
    \KwIn{Global Agent $GA$, Local Agent $LA$, Goal $G_0$}
    
    task\_stack.push($G_0$)
    
    \While{task\_stack not empty}{
        cur\_task $\gets$ task\_stack.pop()
        
        can\_do $\gets$ GA.reflect\_plan(cur\_task)
        
        \If{can\_do}{
            action,obs$\gets$ LA.exec\_task(cur\_task)
            
            cur\_task\_complete $\gets$ GA.reflect\_exec(action,obs) 
        }
        
        \If{\textbf{not} can\_do \textbf{or} \textbf{not} cur\_task\_complete}{
            new\_subtasks $\gets$ GA.plan(cur\_task)
            
            task\_stack.push(new\_subtasks) 
        }
        
        GA.updateMemory()
    }

    \rule{0.9\linewidth}{1pt}
 \caption{Adaptive Planning  of \agent{}}
\label{alg:outline}
\end{algorithm}

\subsection{Multifaceted Memory}

The Memory Module serves as the backbone of \agent's adaptability and learning capabilities, storing historical data to enhance decision-making and reduce redundant actions. It is categorized into five components:

\textbf{Task Memory}: Tracks the execution history of tasks, including task decomposition structures, action traces, success and failure records, and reflections. This hierarchical organization enables efficient retrieval of relevant experiences for task planning and execution.

\textbf{App Memory}: Maintains detailed observations and exploration histories for various applications, including functional descriptions and page-specific interactions. This helps the agent adapt to similar GUI layouts and locate target applications more effectively.

\textbf{Page Memory}: Encompasses the historical steps executed on this interface, such as the positioning of a particular button on the page, among other actions. This facilitates the agent's ability to perform similar operations on the page based on past interactions more effectively.

\textbf{Action Memory}: Incorporates the operations executed during the current task cycle, enabling the agent to more clearly capture the actions performed within this task and to more precisely define the subsequent steps required.

\textbf{User Memory}: Captures user-specific interaction histories, such as preferences, habitual commands, and implicit requirements. This allows \agent{} to better infer user intent and personalize task execution.

\section{Experiments}
To comprehensively compare \agent{} with other GUI agents in handling complex user instructions and executing GUI interactions on mobile devices, we evaluate them using a real-life scenario test set called \bench. Additionally, we assess our method using a widely adopted mobile benchmark, Android Arena.

\subsection{The \bench~Test Set}
\label{sec:mobbench}

The \bench~comprises a diverse test set of 50 tasks designed to evaluate the performance of \agent{} in real-world mobile application scenarios. The test set includes 10 applications widely used in China, each with four tasks of varying difficulty: Easy, Medium, Hard, and Indirect Comprehension, totaling 40 tasks. The tasks are categorized by the complexity and steps required to complete them. Indirect Comprehension is designed for common cases where the user gives a vague instruction without detailing which application or specific steps are required. The agent is expected to decide target application and find an effective approach. Additionally, there are 10 Cross-Application tasks, which involve interacting with two applications and are more close to Hard level in difficulty. These tasks focus on evaluating the ability of information extraction and retrieval, as well as the awareness of sub-goal completion and application switching. 

Compared with several similar task sets mentioned in other papers~\cite{yang2023appagent,wang2024mobile,wang2024mobile2,zhang2024ufo,lee2023explore}, which only get a score when it finishes the task, we assign several milestone scores for sub-tasks in \bench. This allows for a more precise process assessment, in the cases where the task is partially finished. We also include a detailed preparation instruction for tasks when a more reproducible, fair, and stable start is needed. 

To establish a human expert baseline, three human operators independently perform the tasks on three different mobile phones, documenting their execution steps. The average number of steps taken is used as the human expert baseline.

\subsection{Metrics}

Three metrics are designed to better compare the capability of GUI agents thoroughly.

\textbf{Milestone Score (MS)}: Scoring milestones are assigned to several sub-tasks, evenly distributed during the task completion process. Since each task contains 1 to 6 milestones, the agent will get a score as it reaches each milestone. We sum up all milestone scores of 50 tasks as the primary metric.

\textbf{Complete Rate (CR)}: If the agent gets all milestone scores in one task, it is considered as task complete. This is the most common and straightforward metric for GUI agent evaluation.

\textbf{Execution Efficiency (EE)}: We record the effective number of steps for each task and the corresponding milestone scores, that is, the total number of steps executed at the time of getting the last effective milestone score, and calculate the average number of steps required to obtain each effective milestone score. The lower this number, the more efficient the execution; the higher it is, the more it includes ineffective actions.

The average milestone scores and execution steps for each task type are summarized in \Cref{tab:status}.

\begin{table}[h]
\centering

    \small
\begin{tabular}{lcccc}
\toprule
\textbf{Task Type}& \textbf{\# Tasks} & \textbf{\# MS} & \textbf{Avg. Steps} & \textbf{EE}\\
\midrule
Easy & 10 & 10 & 4.3 & 4.30 \\
Medium & 10 & 23 & 7.3 & 3.17 \\
Hard & 10 & 41 & 15.2 & 3.71 \\
Indirect & 10 & 28 & 9.4 & 3.36 \\
Cross-App & 10 & 31 & 10.8 & 3.48 \\
\midrule
Overall & 50 & 133 & 9.4 & 3.53 \\
\bottomrule
\end{tabular}

\caption{\textbf{Milestone scores and expert execution steps for different task types of \bench.}}
\label{tab:status}
\end{table}

\subsection{Setups}
To provide a comprehensive evaluation, \agent{} is compared against several baselines from basic manual operations to several sophisticated agent-based automation.

\textbf{Human Baseline} as mentioned in \Cref{sec:mobbench} are considered as the optimal solution for each task.

\textbf{GPT-4o + Human Baseline} utilizes an iterative process where the GPT model~\cite{gpt4v} provides guidance for manual task execution. 

\textbf{AppAgent}~\cite{yang2023appagent} uses both view hierarchy and screenshot for planning and choosing target actions. All interactive elements are marked with bounding boxes and a unique index for better grounding performance.

\textbf{Mobile Agent (v2)}~\cite{wang2024mobile, wang2024mobile2} uses only visual information from screens as inputs. Target elements are selected with the guidance of OCR and CLIP~\cite{radford2021learning} modules.

\textbf{\agent} is evaluated under several settings by disabling the Memory Module or/and Plan Module to assess its performance and the impact of these two modules. We disable the Plan Module by replacing the Global Agent with a plain agent, and no sub-tasks are provided to the Action and Reflection Module. We disable the Memory Module by removing all in-context examples and historical experience information (including observations, thoughts, previous actions, and their execution status), focusing on assessing the core capability in zero-shot task execution.

All experiments are conducted using \texttt{gpt-4o-2024-05-13} API. The primary evaluation metric is the first attempt complete rate, directly measuring the effectiveness of each system in completing tasks on the first try without retries.

\subsection{Results and Analysis}

The overall experiment results are as listed in \Cref{tab:result}. And for more detailed results categorized by task type please refer to \Cref{fig:result}.

\begin{table}[!ht]
    \small
    \centering
    \setlength\tabcolsep{3.5pt}
    \begin{tabular}{lccc}
        \toprule
        \textbf{Model}                    & \textbf{CR} & \textbf{MS}  & \textbf{EE}    \\
        \midrule
        Human                             & 50/50       & 133          & 3.53           \\
         GPT-4o + Human                    & 49/50       & 131 (98.5\%) & 3.82 (108.2\%) \\ 
        \midrule
        AppAgent                          & 6/50        & 35 (26.3\%)  & 4.43 (125.5\%) \\ 
        MobileAgent (v2)                  & 17/50       & 63 (47.4\%)  & 4.84 (137.1\%) \\ 
        \midrule
        \agent{} w/o M \& P                & 13/50       & 52 (39.1\%)  & 4.42 (125.2\%) \\ 
        \agent{} w/o P                      & 15/50       & 65 (48.9\%)  & 4.17 (118.1\%) \\ 
        \agent{} w/o M                     & 22/50       & 72 (54.1\%)  & 3.81 (107.9\%) \\ 
        \agent{}                            & 28/50       & 88 (66.2\%)  & 3.44 (97.5\%)   \\ 
        \bottomrule
    \end{tabular}
    \caption{\textbf{Overall Performance on \bench.} M: Memory Module. P: Planning Module.}
    \label{tab:result}
\end{table}

\Cref{tab:result} shows the performance of four baselines. Due to the complexity of mobile interfaces and the technical limitations encountered during task execution, the overall task completion rates (Complete Rate, CR) are relatively low for all agents. Consequently, the Milestone Score (MS) serves as a finer metric to more accurately reflect the performance of each agent by considering partial task completion. While there are notable differences in Milestone Scores among the baseline models, the gap in Execution Efficiency (EE) is less significant. This is because most agents can smoothly complete simpler sub-goals, whereas, for more complex sub-goals, the agents either complete them or fail entirely, resulting in closer performance regarding execution efficiency.




\subsubsection{Performance Comparison}

The performance of \textit{MobileAgent} is notably higher than that of \textit{AppAgent}. This improvement is mainly due to the inclusion of both Memory and Reflection modules in \textit{MobileAgent}, which enhance reasoning capacity and utilize more computational resources, such as tokens. Additionally, \textit{MobileAgent} keeps a record of all historical actions, allowing it to learn from the entire sequence of operations, whereas \textit{AppAgent} can only track the most recent action. Furthermore, \textit{MobileAgent} relies on OCR and CLIP modules for target localization, offering greater flexibility and avoiding the technical limitations that \textit{AppAgent} faces when dependent on XML files. By adopting a twice-reflection strategy, the ineffective execution steps are slightly reduced, where the sub-tasks that are not able to be completed with a single action are decomposed finer before executed. This gives clearer guidance for the Local Agent to decide the target actions.

\subsubsection{Ablation Study}

The lower part of  \Cref{tab:result} presents the results of the ablation study, where we experimented with four different configurations by selectively enabling or disabling the Memory and Plan modules. The results indicate that incorporating both Memory and Plan modules significantly enhances the agent's overall performance. 

The Plan module alone shows a much stronger effect than the Memory module alone, validating one of the core contributions of this paper—the effectiveness of task decomposition planning. By decomposing tasks into manageable sub-tasks, \agent{} can perform global planning, avoid redundant actions, and minimize overlooked details, effectively managing its historical actions (since in a tree-structured task, previously completed sub-tasks are inherently tracked). Unlike \textit{MobileAgent}, which focuses solely on the next specific action, \agent{} first determines the next abstract task and then plans the specific execution steps, closely mirroring human reasoning patterns and providing a more structured approach.

When the Memory module is introduced, \agent's performance further improves, particularly in cross-application tasks (see \Cref{fig:result} (b)). This enhancement is due to the Memory module's ability to retain crucial information over longer periods, such as "\texttt{the day I am traveling to Shenzhen}", allowing it to reference previous screens' key content. In contrast, without the Memory module, the agent is limited to short-term memory of only the current and the immediately preceding steps, resulting in less effective task execution.


\subsection{Results on Android Arena}

   \begin{table}[h!]
    \small
    \centering
    \begin{tabular}{ccc}
        \toprule
        Model & SR(single-app) & SR(cross-app) \\
        \midrule
        GPT-3.5 & 0.449 & 0.048 \\
        GPT-4 & 0.759 & 0.571 \\
        \midrule
        \agent(ours) & 0.783 & 0.714 \\
        \bottomrule
    \end{tabular}
    \caption{The performance of LLMs and \agent{} on the Android Arena dataset.}
    \label{fig:aa}
    \end{table} 
    
We also performed evaluations on Android Arena~\cite{xing2024understanding}, comprising 157 single-app tasks and 21 cross-app tasks. As shown in \Cref{fig:aa}, \agent{} achieves success rates (SR) of 0.783 on single-app tasks and 0.714 on cross-app tasks, outperforming GPT-4 by 2.4\% and 14.3\%, respectively. The notable improvement in cross-app tasks is attributed to \agent's subtask decomposition capability, which enables better app-switching decisions during tasks requiring more steps. Additionally, \agent's reflection module encourages exploration, reducing repetitive actions and improving task success rates.

The Android Arena evaluation also highlights limitations in task completion judgment with GPT-4, with 11.8\% of tasks being misclassified, compared to the results checked by humans. This is partly due to \agent's tendency to execute redundant actions after completing tasks, complicating GPT-4's evaluation process. Despite this, \agent's performance gains emphasize its strength in handling complex multi-step tasks, especially in scenarios requiring extensive exploration and app-switching, as evidenced by the significant improvements in cross-app success rates.

\section{Case Study}

\begin{figure*}[!th]
    \centering
    \includegraphics[width=1.0\textwidth,trim=0 0 0 0,clip]{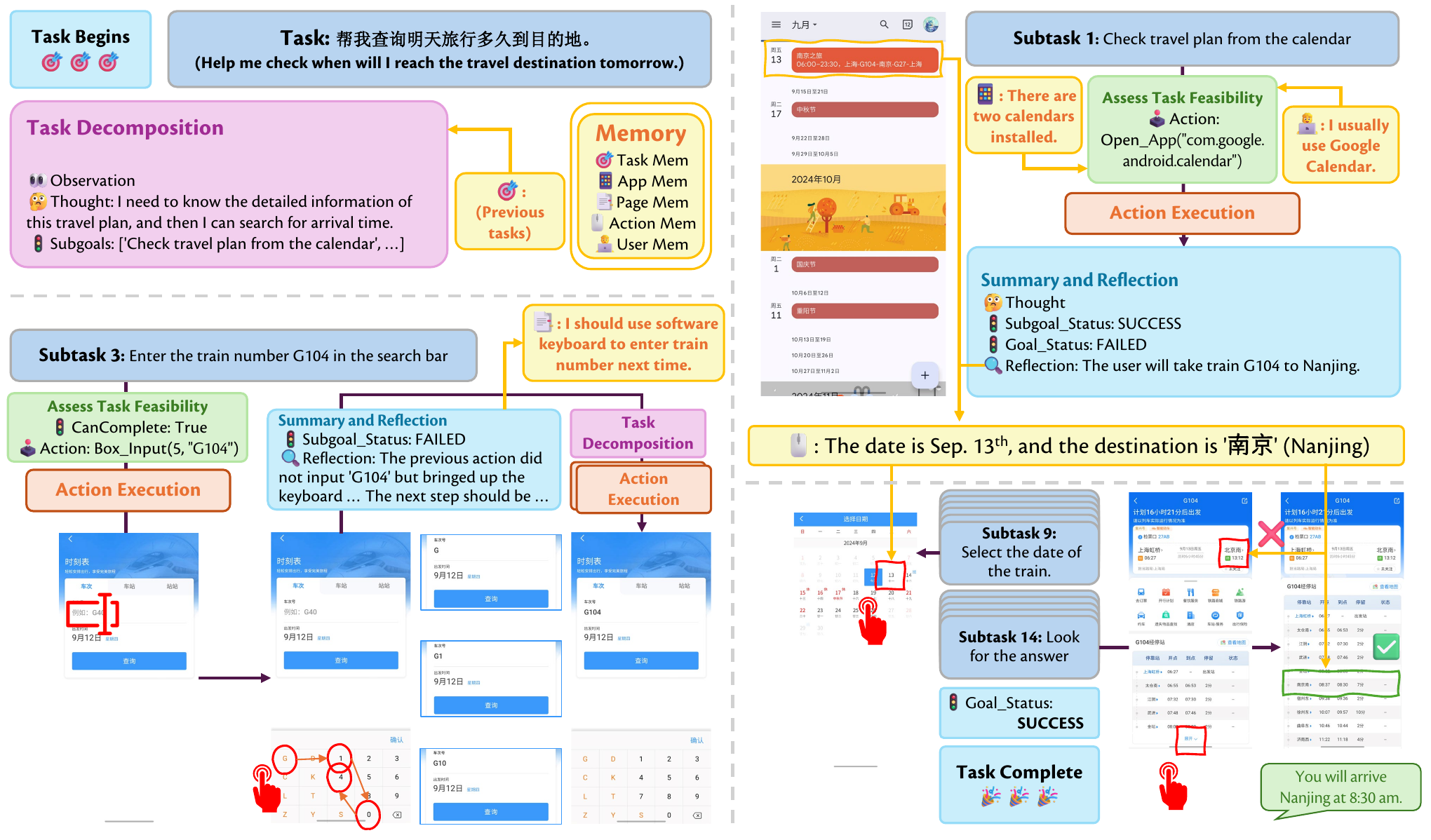}
    \caption{\textbf{The Example Case of \agent.} Please note that several unimportant stages during the execution of a sub-task are omitted for clarity. The key features for each part are as follows. \uline{Task:} \agent{} supports cross-application tasks and can interpret indirect commands.  \uline{Sub-task 1:} Memories are retrieved to select target applications and updated to track the trace. \uline{Sub-task 3:} \agent{} will reflect and try other approaches if the attempt is failed. \uline{Sub-task 9 and sub-task 13:} Memories are used to choose correct actions.}
    \label{fig:case_study_2}
\end{figure*}

\Cref{fig:case_study_2} demonstrates how the adaptive planning and multifaceted memory support task completion in \agent. \agent{} can accurately interpret user intent from command \textit{``Help me check when will I reach the travel destination tomorrow."} and give decomposed sub-tasks based on historical commands. For sub-task 1, \agent{} retrieves relevant details from App and User Memory, extracts key information (train schedule and destination), and stores it in Action Memory. When encountering failures, \agent{} uses historical experiences to reflect and adapt. During sub-task 3, when \agent{} initially failed to input the train number using the \texttt{Box\_Input} function, it reflects on its previous operations and employs a character-by-character input method, completing the task. The key feature of this page will be saved into Page Memory, thus \agent{} is unlikely to encounter the same failure. Additionally, memory retrieval is crucial for handling contextual tasks. In sub-tasks 9 and 13, although the user doesn't explicitly specify the travel date or destination in the task request. \agent{} can rely on previously stored Action Memory data to provide an accurate response.


\section{Related Work}

\subsection{LLM Agents}

The advancements in M/LLMs have significantly influenced the development of agents. LLM-based agents leverage the autonomy, reactivity, proactiveness, and social ability of these models to perceive external environments and make decisions~\cite{xi2023risepotentiallargelanguage}. Emerging abilities, such as CoT reasoning~\cite{NEURIPS2022_9d560961,wang2023selfconsistency,zhang2023automatic} and in-context learning~\cite{NEURIPS2020_1457c0d6,min-etal-2022-rethinking}. Recent studies have explored LLM-based approaches for reflection~\cite{yao2022react,madaan2023selfrefine,shinn2023reflexion,xu2024rejectionimprovesreliabilitytraining}, planning~\cite{sun2023pearl,qian2024investigate,huang2024understandingplanningllmagents}, and memory mechanisms~\cite{zhang2024surveymemorymechanismlarge,zhang2024training,li2023tradinggptmultiagentlayeredmemory,maharana2024evaluatinglongtermconversationalmemory,lan2024depressiondiagnosisdialoguesimulation}. 

At the same time, the agents that utilize M/LLMs to interact with the environments are quickly developed. These agents possess significantly enhanced capabilities for environment observation, task decomposition, and action decision-making, which enable M/LLMs to solve complex tasks across social simulations~\cite{10.1145/3586183.3606763,10.5555/3618408.3618425,10.1145/3544548.3581503,lan2024depressiondiagnosisdialoguesimulation}, embodied robots~\cite{wu2023embodiedtaskplanninglarge}, software development~\cite{qian-etal-2024-chatdev,qian-etal-2024-experiential} and virtual assistants~\cite{wang2023voyageropenendedembodiedagent}.

\subsection{GUI Agents}

\subsubsection{Traditional GUI Agents} Controlling GUI screens based on user commands is a complex task that involves both GUI understanding and command interpretation. Early approaches to GUI agents focused on embedding and modular systems. For example, several agents~\cite{li2017sugilite,li2019pumice} combined natural language and programming demonstrations, allowing users to define tasks via descriptions and demonstrations. This method relied on text and image matching for script-based control of the interface.  Traditional GUI agents were largely limited by their reliance on pre-defined rules and manual programming. These agents were effective within controlled environments but struggled with dynamic, real-world GUI contexts due to their lack of flexibility and adaptability. They required specific scripts or rules for each task, making them less robust when handling the diverse and evolving nature of real-world applications.

\subsubsection{Advancements with Multimodal Pretrain Models} The advent of multimodal pretraining models~\cite{bai2021uibertlearninggenericmultimodal,li2021vutversatileuitransformer,lispotlight,he2021actionbert,li2021screen2vec,wang2021screen2words,fu2024understanding} for GUI understanding marked a significant shift in the development of GUI agents. Pretrained agents~\cite{sun2022meta,zhu2023cam,zhan2023autoui,10.1145/3616855.3635753} integrated multimodal information, such as dialogue history, screenshots, and action history, through pretraining. Unlike earlier methods that relied on rigid scripts, these end-to-end models adopted a more human-like approach to interacting with interfaces, enhancing their efficiency in information retrieval and task execution by mapping visual observations and text commands directly into actions.

\subsubsection{MLLM-Empowered GUI Agents} The integration of MLLMs in GUI agents has introduced new opportunities to further enhance their capabilities. With the rise of larger scale models, GUI agents~\cite{yang2023appagent,zhang2024ufo,lee2023explore} began to leverage advanced reasoning and decision-making processes. These models utilized structural information provided in the view hierarchy (VH) to annotate and locate UI elements, guiding a sequence of atomic actions to achieve specific goals. VH-only agents~\cite{wen2024autodroid} depend on the structural information to reason and make decisions, which greatly lowers the cost of inference making it suitable for deployment on the device. Image-only agents~\cite{wang2024mobile,wang2024mobile2,gao2024assistgui,yan2023gpt4vwonderlandlargemultimodal}, which employs OCR, CLIP~\cite{radford2021learning} module, and object detection methods to identify operation targets. This image-only approach is particularly effective when the view hierarchy is inaccessible or noisy, but it may also encounter challenges, e.g. opening a target application by clicking when names are hidden, or logos vary across screens.

\section{Conclusion and Future Works}

This paper presented \agent, an innovative \textbf{Mob}ile phone \textbf{A}ssistant system empowered by MLLMs. Utilizing a two-level agent structure, comprising a Global Agent and a Local Agent, \agent{} effectively understands user commands, plans tasks, and executes actions. The combination of Memory and Plan Modules enhances its ability to learn from previous interactions, improving efficiency and accuracy. Our evaluations demonstrated that \agent~surpasses existing mobile assistants in handling complex tasks, leveraging multi-level memory, task decomposition, and action-validation mechanisms. These features enable precise task execution even with intricate or indirect commands. Future work will focus on improving the performance on image-only scenarios where the view hierarchy is unattainable, deploying an end-side model on mobile phones for faster response and secured privacy. We will continue to expand \bench{} by adding more popular applications from different regions and languages.
We hope \agent{} illustrates the potential of MLLMs-empowered mobile assistants and provides valuable insights for future works.

\section*{Acknowledgment}
This work is funded by the China NSFC Projects (92370206, 62120106006, U23B2057) and Shanghai Municipal Science and Technology Major Project (2021SHZDZX0102).

\newpage
\bibliography{custom}

\begin{thebibliography}{76}
\providecommand{\natexlab}[1]{#1}

\bibitem[{Aher et~al.(2023)Aher, Arriaga, and Kalai}]{10.5555/3618408.3618425}
Gati Aher, Rosa~I. Arriaga, and Adam~Tauman Kalai. 2023.
\newblock Using large language models to simulate multiple humans and replicate human subject studies.
\newblock In \emph{Proceedings of the 40th International Conference on Machine Learning}, ICML'23. JMLR.org.

\bibitem[{Bai et~al.(2021)Bai, Zang, Xu, Sunkara, Rastogi, Chen, and y~Arcas}]{bai2021uibertlearninggenericmultimodal}
Chongyang Bai, Xiaoxue Zang, Ying Xu, Srinivas Sunkara, Abhinav Rastogi, Jindong Chen, and Blaise~Aguera y~Arcas. 2021.
\newblock \href {https://arxiv.org/abs/2107.13731} {Uibert: Learning generic multimodal representations for ui understanding}.
\newblock \emph{Preprint}, arXiv:2107.13731.

\bibitem[{Brown et~al.(2020)Brown, Mann, Ryder, Subbiah, Kaplan, Dhariwal, Neelakantan, Shyam, Sastry, Askell, Agarwal, Herbert-Voss, Krueger, Henighan, Child, Ramesh, Ziegler, Wu, Winter, Hesse, Chen, Sigler, Litwin, Gray, Chess, Clark, Berner, McCandlish, Radford, Sutskever, and Amodei}]{NEURIPS2020_1457c0d6}
Tom Brown, Benjamin Mann, Nick Ryder, Melanie Subbiah, Jared~D Kaplan, Prafulla Dhariwal, Arvind Neelakantan, Pranav Shyam, Girish Sastry, Amanda Askell, Sandhini Agarwal, Ariel Herbert-Voss, Gretchen Krueger, Tom Henighan, Rewon Child, Aditya Ramesh, Daniel Ziegler, Jeffrey Wu, Clemens Winter, Chris Hesse, Mark Chen, Eric Sigler, Mateusz Litwin, Scott Gray, Benjamin Chess, Jack Clark, Christopher Berner, Sam McCandlish, Alec Radford, Ilya Sutskever, and Dario Amodei. 2020.
\newblock \href {https://proceedings.neurips.cc/paper_files/paper/2020/file/1457c0d6bfcb4967418bfb8ac142f64a-Paper.pdf} {Language models are few-shot learners}.
\newblock In \emph{Advances in Neural Information Processing Systems}, volume~33, pages 1877--1901. Curran Associates, Inc.

\bibitem[{Cao et~al.(2024)Cao, Lei, Wu, Chen, Fu, Gao, Xiong, Zhang, Hu, Mao, Xie, Xu, Zhang, Wang, Sun, Yin, Xiong, Ni, Liu, Zhong, Chen, Yu, and Yu}]{cao2024spider2vfarmultimodalagents}
Ruisheng Cao, Fangyu Lei, Haoyuan Wu, Jixuan Chen, Yeqiao Fu, Hongcheng Gao, Xinzhuang Xiong, Hanchong Zhang, Wenjing Hu, Yuchen Mao, Tianbao Xie, Hongshen Xu, Danyang Zhang, Sida Wang, Ruoxi Sun, Pengcheng Yin, Caiming Xiong, Ansong Ni, Qian Liu, Victor Zhong, Lu~Chen, Kai Yu, and Tao Yu. 2024.
\newblock \href {https://proceedings.neurips.cc/paper_files/paper/2024/file/c2f71567cd53464161cab3336e8fc865-Paper-Datasets_and_Benchmarks_Track.pdf} {Spider2-v: How far are multimodal agents from automating data science and engineering workflows?}

\bibitem[{Chen et~al.(2023)Chen, Zhu, Shen, Li, Liu, Zhang, Krishnamoorthi, Chandra, Xiong, and Elhoseiny}]{chen2023minigptv2largelanguagemodel}
Jun Chen, Deyao Zhu, Xiaoqian Shen, Xiang Li, Zechun Liu, Pengchuan Zhang, Raghuraman Krishnamoorthi, Vikas Chandra, Yunyang Xiong, and Mohamed Elhoseiny. 2023.
\newblock \href {https://arxiv.org/abs/2310.09478} {Minigpt-v2: large language model as a unified interface for vision-language multi-task learning}.
\newblock \emph{Preprint}, arXiv:2310.09478.

\bibitem[{Chen et~al.(2024{\natexlab{a}})Chen, Xu, Qi, and Guo}]{chen2024mllmstrongrerankeradvancing}
Zhanpeng Chen, Chengjin Xu, Yiyan Qi, and Jian Guo. 2024{\natexlab{a}}.
\newblock \href {https://arxiv.org/abs/2407.21439} {Mllm is a strong reranker: Advancing multimodal retrieval-augmented generation via knowledge-enhanced reranking and noise-injected training}.
\newblock \emph{Preprint}, arXiv:2407.21439.

\bibitem[{Chen et~al.(2024{\natexlab{b}})Chen, Wu, Wang, Su, Chen, Xing, Zhong, Zhang, Zhu, Lu, Li, Luo, Lu, Qiao, and Dai}]{chen2024internvlscalingvisionfoundation}
Zhe Chen, Jiannan Wu, Wenhai Wang, Weijie Su, Guo Chen, Sen Xing, Muyan Zhong, Qinglong Zhang, Xizhou Zhu, Lewei Lu, Bin Li, Ping Luo, Tong Lu, Yu~Qiao, and Jifeng Dai. 2024{\natexlab{b}}.
\newblock \href {https://doi.org/10.1109/CVPR52733.2024.02283} {Internvl: Scaling up vision foundation models and aligning for generic visual-linguistic tasks}.
\newblock In \emph{2024 IEEE/CVF Conference on Computer Vision and Pattern Recognition (CVPR)}, pages 24185--24198.

\bibitem[{Dai et~al.(2023)Dai, Li, LI, Tiong, Zhao, Wang, Li, Fung, and Hoi}]{NEURIPS2023_9a6a435e}
Wenliang Dai, Junnan Li, DONGXU LI, Anthony Tiong, Junqi Zhao, Weisheng Wang, Boyang Li, Pascale~N Fung, and Steven Hoi. 2023.
\newblock \href {https://proceedings.neurips.cc/paper_files/paper/2023/file/9a6a435e75419a836fe47ab6793623e6-Paper-Conference.pdf} {Instructblip: Towards general-purpose vision-language models with instruction tuning}.
\newblock In \emph{Advances in Neural Information Processing Systems}, volume~36, pages 49250--49267. Curran Associates, Inc.

\bibitem[{Fu et~al.(2024)Fu, Zhang, Wang, Zeng, and Zheng}]{fu2024understanding}
Jingwen Fu, Xiaoyi Zhang, Yuwang Wang, Wenjun Zeng, and Nanning Zheng. 2024.
\newblock Understanding mobile gui: From pixel-words to screen-sentences.
\newblock \emph{Neurocomputing}, 601:128200.

\bibitem[{Gao et~al.(2024)Gao, Ji, Bai, Ouyang, Li, Mao, Wu, Zhang, Wang, Guo et~al.}]{gao2024assistgui}
Difei Gao, Lei Ji, Zechen Bai, Mingyu Ouyang, Peiran Li, Dongxing Mao, Qinchen Wu, Weichen Zhang, Peiyi Wang, Xiangwu Guo, et~al. 2024.
\newblock Assistgui: Task-oriented pc graphical user interface automation.
\newblock In \emph{Proceedings of the IEEE/CVF Conference on Computer Vision and Pattern Recognition}, pages 13289--13298.

\bibitem[{Ge et~al.(2024)Ge, Zhao, Li, Ge, and Shan}]{ge2024seeddataedittechnicalreporthybrid}
Yuying Ge, Sijie Zhao, Chen Li, Yixiao Ge, and Ying Shan. 2024.
\newblock \href {https://arxiv.org/abs/2405.04007} {Seed-data-edit technical report: A hybrid dataset for instructional image editing}.
\newblock \emph{Preprint}, arXiv:2405.04007.

\bibitem[{He et~al.(2021)He, Sunkara, Zang, Xu, Liu, Wichers, Schubiner, Lee, and Chen}]{he2021actionbert}
Zecheng He, Srinivas Sunkara, Xiaoxue Zang, Ying Xu, Lijuan Liu, Nevan Wichers, Gabriel Schubiner, Ruby Lee, and Jindong Chen. 2021.
\newblock Actionbert: Leveraging user actions for semantic understanding of user interfaces.
\newblock In \emph{Proceedings of the AAAI Conference on Artificial Intelligence}, volume~35, pages 5931--5938.

\bibitem[{Hu et~al.(2024)Hu, Yao, Wang, WANG, Pan, Chen, Yu, Wu, Zhao, Zhang, Han, Lin, Xue, dahai li, Liu, and Sun}]{hu2023viscpm}
Jinyi Hu, Yuan Yao, Chongyi Wang, SHAN WANG, Yinxu Pan, Qianyu Chen, Tianyu Yu, Hanghao Wu, Yue Zhao, Haoye Zhang, Xu~Han, Yankai Lin, Jiao Xue, dahai li, Zhiyuan Liu, and Maosong Sun. 2024.
\newblock \href {https://openreview.net/forum?id=Kuh5qgCGCp} {Large multilingual models pivot zero-shot multimodal learning across languages}.
\newblock In \emph{The Twelfth International Conference on Learning Representations}.

\bibitem[{Huang et~al.(2024)Huang, Liu, Chen, Wang, Wang, Lian, Wang, Tang, and Chen}]{huang2024understandingplanningllmagents}
Xu~Huang, Weiwen Liu, Xiaolong Chen, Xingmei Wang, Hao Wang, Defu Lian, Yasheng Wang, Ruiming Tang, and Enhong Chen. 2024.
\newblock \href {https://arxiv.org/abs/2402.02716} {Understanding the planning of llm agents: A survey}.
\newblock \emph{Preprint}, arXiv:2402.02716.

\bibitem[{Jo et~al.(2023)Jo, Epstein, Jung, and Kim}]{10.1145/3544548.3581503}
Eunkyung Jo, Daniel~A. Epstein, Hyunhoon Jung, and Young-Ho Kim. 2023.
\newblock \href {https://doi.org/10.1145/3544548.3581503} {Understanding the benefits and challenges of deploying conversational ai leveraging large language models for public health intervention}.
\newblock In \emph{Proceedings of the 2023 CHI Conference on Human Factors in Computing Systems}, CHI '23, New York, NY, USA. Association for Computing Machinery.

\bibitem[{Lan et~al.(2024)Lan, Jin, Zhu, Chen, Zhang, Zhu, and Wu}]{lan2024depressiondiagnosisdialoguesimulation}
Kunyao Lan, Bingui Jin, Zichen Zhu, Siyuan Chen, Shu Zhang, Kenny~Q. Zhu, and Mengyue Wu. 2024.
\newblock \href {https://arxiv.org/abs/2409.15084} {Depression diagnosis dialogue simulation: Self-improving psychiatrist with tertiary memory}.
\newblock \emph{Preprint}, arXiv:2409.15084.

\bibitem[{Lee et~al.(2024{\natexlab{a}})Lee, Choi, Lee, Wasi, Choi, Ko, Oh, and Shin}]{lee2023explore}
Sunjae Lee, Junyoung Choi, Jungjae Lee, Munim~Hasan Wasi, Hojun Choi, Steven~Y. Ko, Sangeun Oh, and Insik Shin. 2024{\natexlab{a}}.
\newblock \href {https://arxiv.org/abs/2312.03003} {Explore, select, derive, and recall: Augmenting llm with human-like memory for mobile task automation}.
\newblock \emph{Preprint}, arXiv:2312.03003.

\bibitem[{Lee et~al.(2024{\natexlab{b}})Lee, Jeon, Lee, Byun, Son, Shin, Ko, and Kim}]{lee2024llavadocentinstructiontuningmultimodal}
Unggi Lee, Minji Jeon, Yunseo Lee, Gyuri Byun, Yoorim Son, Jaeyoon Shin, Hongkyu Ko, and Hyeoncheol Kim. 2024{\natexlab{b}}.
\newblock \href {https://arxiv.org/abs/2402.06264} {Llava-docent: Instruction tuning with multimodal large language model to support art appreciation education}.
\newblock \emph{Preprint}, arXiv:2402.06264.

\bibitem[{Li and Li(2023)}]{lispotlight}
Gang Li and Yang Li. 2023.
\newblock Spotlight: Mobile ui understanding using vision-language models with a focus.
\newblock In \emph{The Eleventh International Conference on Learning Representations}.

\bibitem[{Li et~al.(2017)Li, Azaria, and Myers}]{li2017sugilite}
Toby Jia-Jun Li, Amos Azaria, and Brad~A. Myers. 2017.
\newblock \href {https://doi.org/10.1145/3025453.3025483} {Sugilite: Creating multimodal smartphone automation by demonstration}.
\newblock In \emph{Proceedings of the 2017 CHI Conference on Human Factors in Computing Systems}, CHI '17, page 6038–6049, New York, NY, USA. Association for Computing Machinery.

\bibitem[{Li et~al.(2021{\natexlab{a}})Li, Popowski, Mitchell, and Myers}]{li2021screen2vec}
Toby Jia-Jun Li, Lindsay Popowski, Tom Mitchell, and Brad~A Myers. 2021{\natexlab{a}}.
\newblock Screen2vec: Semantic embedding of gui screens and gui components.
\newblock In \emph{Proceedings of the 2021 CHI Conference on Human Factors in Computing Systems}, pages 1--15.

\bibitem[{Li et~al.(2019)Li, Radensky, Jia, Singarajah, Mitchell, and Myers}]{li2019pumice}
Toby Jia-Jun Li, Marissa Radensky, Justin Jia, Kirielle Singarajah, Tom~M Mitchell, and Brad~A Myers. 2019.
\newblock Pumice: A multi-modal agent that learns concepts and conditionals from natural language and demonstrations.
\newblock In \emph{Proceedings of the 32nd annual ACM symposium on user interface software and technology}, pages 577--589.

\bibitem[{Li et~al.(2021{\natexlab{b}})Li, Li, Zhou, Dehghani, and Gritsenko}]{li2021vutversatileuitransformer}
Yang Li, Gang Li, Xin Zhou, Mostafa Dehghani, and Alexey Gritsenko. 2021{\natexlab{b}}.
\newblock \href {https://arxiv.org/abs/2112.05692} {Vut: Versatile ui transformer for multi-modal multi-task user interface modeling}.
\newblock \emph{Preprint}, arXiv:2112.05692.

\bibitem[{Li et~al.(2023)Li, Yu, Li, Chen, and Khashanah}]{li2023tradinggptmultiagentlayeredmemory}
Yang Li, Yangyang Yu, Haohang Li, Zhi Chen, and Khaldoun Khashanah. 2023.
\newblock \href {https://arxiv.org/abs/2309.03736} {Tradinggpt: Multi-agent system with layered memory and distinct characters for enhanced financial trading performance}.
\newblock \emph{Preprint}, arXiv:2309.03736.

\bibitem[{Liu et~al.(2024{\natexlab{a}})Liu, Li, Li, and Lee}]{liu2024improved}
Haotian Liu, Chunyuan Li, Yuheng Li, and Yong~Jae Lee. 2024{\natexlab{a}}.
\newblock Improved baselines with visual instruction tuning.
\newblock In \emph{Proceedings of the IEEE/CVF Conference on Computer Vision and Pattern Recognition (CVPR)}, pages 26296--26306.

\bibitem[{Liu et~al.(2023)Liu, Li, Wu, and Lee}]{NEURIPS2023_6dcf277e}
Haotian Liu, Chunyuan Li, Qingyang Wu, and Yong~Jae Lee. 2023.
\newblock \href {https://proceedings.neurips.cc/paper_files/paper/2023/file/6dcf277ea32ce3288914faf369fe6de0-Paper-Conference.pdf} {Visual instruction tuning}.
\newblock In \emph{Advances in Neural Information Processing Systems}, volume~36, pages 34892--34916. Curran Associates, Inc.

\bibitem[{Liu et~al.(2024{\natexlab{b}})Liu, He, Guo, Li, Jin, Li, Li, Chan, Chen, Xue, Luo, Liu, and Guo}]{liu2024hiprompttuningfreehigherresolutiongeneration}
Xinyu Liu, Yingqing He, Lanqing Guo, Xiang Li, Bu~Jin, Peng Li, Yan Li, Chi-Min Chan, Qifeng Chen, Wei Xue, Wenhan Luo, Qifeng Liu, and Yike Guo. 2024{\natexlab{b}}.
\newblock \href {https://arxiv.org/abs/2409.02919} {Hiprompt: Tuning-free higher-resolution generation with hierarchical mllm prompts}.
\newblock \emph{Preprint}, arXiv:2409.02919.

\bibitem[{Ma et~al.(2024)Ma, Zhang, and Zhao}]{ma-etal-2024-coco}
Xinbei Ma, Zhuosheng Zhang, and Hai Zhao. 2024.
\newblock \href {https://aclanthology.org/2024.findings-acl.539} {{C}o{C}o-agent: A comprehensive cognitive {MLLM} agent for smartphone {GUI} automation}.
\newblock In \emph{Findings of the Association for Computational Linguistics ACL 2024}, pages 9097--9110, Bangkok, Thailand and virtual meeting. Association for Computational Linguistics.

\bibitem[{Madaan et~al.(2023)Madaan, Tandon, Gupta, Hallinan, Gao, Wiegreffe, Alon, Dziri, Prabhumoye, Yang et~al.}]{madaan2023selfrefine}
Aman Madaan, Niket Tandon, Prakhar Gupta, Skyler Hallinan, Luyu Gao, Sarah Wiegreffe, Uri Alon, Nouha Dziri, Shrimai Prabhumoye, Yiming Yang, et~al. 2023.
\newblock Self-refine: Iterative refinement with self-feedback.

\bibitem[{Maharana et~al.(2024)Maharana, Lee, Tulyakov, Bansal, Barbieri, and Fang}]{maharana2024evaluatinglongtermconversationalmemory}
Adyasha Maharana, Dong-Ho Lee, Sergey Tulyakov, Mohit Bansal, Francesco Barbieri, and Yuwei Fang. 2024.
\newblock Evaluating very long-term conversational memory of llm agents.

\bibitem[{Min et~al.(2022)Min, Lyu, Holtzman, Artetxe, Lewis, Hajishirzi, and Zettlemoyer}]{min-etal-2022-rethinking}
Sewon Min, Xinxi Lyu, Ari Holtzman, Mikel Artetxe, Mike Lewis, Hannaneh Hajishirzi, and Luke Zettlemoyer. 2022.
\newblock \href {https://doi.org/10.18653/v1/2022.emnlp-main.759} {Rethinking the role of demonstrations: What makes in-context learning work?}
\newblock In \emph{Proceedings of the 2022 Conference on Empirical Methods in Natural Language Processing}, pages 11048--11064, Abu Dhabi, United Arab Emirates. Association for Computational Linguistics.

\bibitem[{Nong et~al.(2024)Nong, Zhu, Wu, Jin, Shan, Huang, and Xu}]{nong2024mobileflowmultimodalllmmobile}
Songqin Nong, Jiali Zhu, Rui Wu, Jiongchao Jin, Shuo Shan, Xiutian Huang, and Wenhao Xu. 2024.
\newblock \href {https://arxiv.org/abs/2407.04346} {Mobileflow: A multimodal llm for mobile gui agent}.
\newblock \emph{Preprint}, arXiv:2407.04346.

\bibitem[{OpenAI(2023)}]{gpt4v}
OpenAI. 2023.
\newblock Gpt-4v(ision) system card.
\newblock \url{https://openai.com/research/gpt-4v-system-card}.

\bibitem[{Pan et~al.(2024)Pan, Dong, Huang, Peng, Chen, and Wei}]{pan2024kosmosg}
Xichen Pan, Li~Dong, Shaohan Huang, Zhiliang Peng, Wenhu Chen, and Furu Wei. 2024.
\newblock \href {https://openreview.net/forum?id=he6mX9LTyE} {Kosmos-g: Generating images in context with multimodal large language models}.
\newblock In \emph{The Twelfth International Conference on Learning Representations}.

\bibitem[{Park et~al.(2023)Park, O'Brien, Cai, Morris, Liang, and Bernstein}]{10.1145/3586183.3606763}
Joon~Sung Park, Joseph O'Brien, Carrie~Jun Cai, Meredith~Ringel Morris, Percy Liang, and Michael~S. Bernstein. 2023.
\newblock \href {https://doi.org/10.1145/3586183.3606763} {Generative agents: Interactive simulacra of human behavior}.
\newblock In \emph{Proceedings of the 36th Annual ACM Symposium on User Interface Software and Technology}, UIST '23, New York, NY, USA. Association for Computing Machinery.

\bibitem[{Qian et~al.(2024{\natexlab{a}})Qian, Dang, Li, Liu, Xie, Wang, Chen, Yang, Cong, Che, Liu, and Sun}]{qian-etal-2024-experiential}
Chen Qian, Yufan Dang, Jiahao Li, Wei Liu, Zihao Xie, YiFei Wang, Weize Chen, Cheng Yang, Xin Cong, Xiaoyin Che, Zhiyuan Liu, and Maosong Sun. 2024{\natexlab{a}}.
\newblock \href {https://aclanthology.org/2024.acl-long.305} {Experiential co-learning of software-developing agents}.
\newblock In \emph{Proceedings of the 62nd Annual Meeting of the Association for Computational Linguistics (Volume 1: Long Papers)}, pages 5628--5640, Bangkok, Thailand. Association for Computational Linguistics.

\bibitem[{Qian et~al.(2024{\natexlab{b}})Qian, Liu, Liu, Chen, Dang, Li, Yang, Chen, Su, Cong, Xu, Li, Liu, and Sun}]{qian-etal-2024-chatdev}
Chen Qian, Wei Liu, Hongzhang Liu, Nuo Chen, Yufan Dang, Jiahao Li, Cheng Yang, Weize Chen, Yusheng Su, Xin Cong, Juyuan Xu, Dahai Li, Zhiyuan Liu, and Maosong Sun. 2024{\natexlab{b}}.
\newblock \href {https://aclanthology.org/2024.acl-long.810} {{C}hat{D}ev: Communicative agents for software development}.
\newblock In \emph{Proceedings of the 62nd Annual Meeting of the Association for Computational Linguistics (Volume 1: Long Papers)}, pages 15174--15186, Bangkok, Thailand. Association for Computational Linguistics.

\bibitem[{Qian et~al.(2024{\natexlab{c}})Qian, Liang, Qin, Ye, Cong, Lin, Wu, Liu, and Sun}]{qian2024investigate}
Cheng Qian, Shihao Liang, Yujia Qin, Yining Ye, Xin Cong, Yankai Lin, Yesai Wu, Zhiyuan Liu, and Maosong Sun. 2024{\natexlab{c}}.
\newblock \href {https://arxiv.org/abs/2401.13996} {Investigate-consolidate-exploit: A general strategy for inter-task agent self-evolution}.
\newblock \emph{Preprint}, arXiv:2401.13996.

\bibitem[{Radford et~al.(2021)Radford, Kim, Hallacy, Ramesh, Goh, Agarwal, Sastry, Askell, Mishkin, Clark, Krueger, and Sutskever}]{radford2021learning}
Alec Radford, Jong~Wook Kim, Chris Hallacy, Aditya Ramesh, Gabriel Goh, Sandhini Agarwal, Girish Sastry, Amanda Askell, Pamela Mishkin, Jack Clark, Gretchen Krueger, and Ilya Sutskever. 2021.
\newblock \href {https://proceedings.mlr.press/v139/radford21a.html} {Learning transferable visual models from natural language supervision}.
\newblock In \emph{Proceedings of the 38th International Conference on Machine Learning}, volume 139 of \emph{Proceedings of Machine Learning Research}, pages 8748--8763. PMLR.

\bibitem[{Shinn et~al.(2023)Shinn, Cassano, Gopinath, Narasimhan, and Yao}]{shinn2023reflexion}
Noah Shinn, Federico Cassano, Ashwin Gopinath, Karthik Narasimhan, and Shunyu Yao. 2023.
\newblock \href {https://proceedings.neurips.cc/paper_files/paper/2023/file/1b44b878bb782e6954cd888628510e90-Paper-Conference.pdf} {Reflexion: language agents with verbal reinforcement learning}.
\newblock In \emph{Advances in Neural Information Processing Systems}, volume~36, pages 8634--8652. Curran Associates, Inc.

\bibitem[{Sun et~al.(2022)Sun, Chen, Chen, Dai, Zhu, and Yu}]{sun2022meta}
Liangtai Sun, Xingyu Chen, Lu~Chen, Tianle Dai, Zichen Zhu, and Kai Yu. 2022.
\newblock \href {https://doi.org/10.18653/v1/2022.emnlp-main.449} {{META}-{GUI}: Towards multi-modal conversational agents on mobile {GUI}}.
\newblock In \emph{Proceedings of the 2022 Conference on Empirical Methods in Natural Language Processing}, pages 6699--6712, Abu Dhabi, United Arab Emirates. Association for Computational Linguistics.

\bibitem[{Sun et~al.(2024{\natexlab{a}})Sun, Cui, Zhang, Zhang, Yu, Wang, Rao, Liu, Huang, and Wang}]{sun2024generativemultimodalmodelsincontext}
Quan Sun, Yufeng Cui, Xiaosong Zhang, Fan Zhang, Qiying Yu, Yueze Wang, Yongming Rao, Jingjing Liu, Tiejun Huang, and Xinlong Wang. 2024{\natexlab{a}}.
\newblock Generative multimodal models are in-context learners.

\bibitem[{Sun et~al.(2024{\natexlab{b}})Sun, Liu, Wang, Iter, Zhu, and Iyyer}]{sun2023pearl}
Simeng Sun, Yang Liu, Shuohang Wang, Dan Iter, Chenguang Zhu, and Mohit Iyyer. 2024{\natexlab{b}}.
\newblock \href {https://aclanthology.org/2024.eacl-long.29} {{PEARL}: Prompting large language models to plan and execute actions over long documents}.
\newblock In \emph{Proceedings of the 18th Conference of the European Chapter of the Association for Computational Linguistics (Volume 1: Long Papers)}, pages 469--486, St. Julian{'}s, Malta. Association for Computational Linguistics.

\bibitem[{Team(2024)}]{team2023gemini}
Gemini Team. 2024.
\newblock \href {https://arxiv.org/abs/2312.11805} {Gemini: A family of highly capable multimodal models}.
\newblock \emph{Preprint}, arXiv:2312.11805.

\bibitem[{Wang et~al.(2021)Wang, Li, Zhou, Chen, Grossman, and Li}]{wang2021screen2words}
Bryan Wang, Gang Li, Xin Zhou, Zhourong Chen, Tovi Grossman, and Yang Li. 2021.
\newblock \href {https://doi.org/10.1145/3472749.3474765} {Screen2words: Automatic mobile ui summarization with multimodal learning}.
\newblock In \emph{The 34th Annual ACM Symposium on User Interface Software and Technology}, UIST '21, page 498–510, New York, NY, USA. Association for Computing Machinery.

\bibitem[{Wang et~al.(2023{\natexlab{a}})Wang, Xie, Jiang, Mandlekar, Xiao, Zhu, Fan, and Anandkumar}]{wang2023voyageropenendedembodiedagent}
Guanzhi Wang, Yuqi Xie, Yunfan Jiang, Ajay Mandlekar, Chaowei Xiao, Yuke Zhu, Linxi Fan, and Anima Anandkumar. 2023{\natexlab{a}}.
\newblock \href {https://arxiv.org/abs/2305.16291} {Voyager: An open-ended embodied agent with large language models}.
\newblock \emph{Preprint}, arXiv:2305.16291.

\bibitem[{Wang et~al.(2025)Wang, Xu, Jia, Zhang, Yan, Shen, Zhang, Huang, and Sang}]{wang2024mobile2}
Junyang Wang, Haiyang Xu, Haitao Jia, Xi~Zhang, Ming Yan, Weizhou Shen, Ji~Zhang, Fei Huang, and Jitao Sang. 2025.
\newblock Mobile-agent-v2: Mobile device operation assistant with effective navigation via multi-agent collaboration.
\newblock \emph{Advances in Neural Information Processing Systems}, 37:2686--2710.

\bibitem[{Wang et~al.(2024{\natexlab{a}})Wang, Xu, Ye, Yan, Shen, Zhang, Huang, and Sang}]{wang2024mobile}
Junyang Wang, Haiyang Xu, Jiabo Ye, Ming Yan, Weizhou Shen, Ji~Zhang, Fei Huang, and Jitao Sang. 2024{\natexlab{a}}.
\newblock \href {https://arxiv.org/abs/2401.16158} {Mobile-agent: Autonomous multi-modal mobile device agent with visual perception}.
\newblock \emph{Preprint}, arXiv:2401.16158.

\bibitem[{Wang et~al.(2023{\natexlab{b}})Wang, Wei, Schuurmans, Le, Chi, Narang, Chowdhery, and Zhou}]{wang2023selfconsistency}
Xuezhi Wang, Jason Wei, Dale Schuurmans, Quoc~V Le, Ed~H. Chi, Sharan Narang, Aakanksha Chowdhery, and Denny Zhou. 2023{\natexlab{b}}.
\newblock \href {https://openreview.net/forum?id=1PL1NIMMrw} {Self-consistency improves chain of thought reasoning in language models}.
\newblock In \emph{The Eleventh International Conference on Learning Representations}.

\bibitem[{Wang et~al.(2024{\natexlab{b}})Wang, Chen, Han, Lin, Zhao, Liu, Zhai, Yuan, You, and Yang}]{wang2024exploringreasoningabilitiesmultimodal}
Yiqi Wang, Wentao Chen, Xiaotian Han, Xudong Lin, Haiteng Zhao, Yongfei Liu, Bohan Zhai, Jianbo Yuan, Quanzeng You, and Hongxia Yang. 2024{\natexlab{b}}.
\newblock \href {https://arxiv.org/abs/2401.06805} {Exploring the reasoning abilities of multimodal large language models (mllms): A comprehensive survey on emerging trends in multimodal reasoning}.
\newblock \emph{Preprint}, arXiv:2401.06805.

\bibitem[{Wang and Zhao(2023)}]{wang2023geminireasoningunveilingcommonsense}
Yuqing Wang and Yun Zhao. 2023.
\newblock \href {https://arxiv.org/abs/2312.17661} {Gemini in reasoning: Unveiling commonsense in multimodal large language models}.
\newblock \emph{Preprint}, arXiv:2312.17661.

\bibitem[{Wei et~al.(2022)Wei, Wang, Schuurmans, Bosma, ichter, Xia, Chi, Le, and Zhou}]{NEURIPS2022_9d560961}
Jason Wei, Xuezhi Wang, Dale Schuurmans, Maarten Bosma, brian ichter, Fei Xia, Ed~Chi, Quoc~V Le, and Denny Zhou. 2022.
\newblock \href {https://proceedings.neurips.cc/paper_files/paper/2022/file/9d5609613524ecf4f15af0f7b31abca4-Paper-Conference.pdf} {Chain-of-thought prompting elicits reasoning in large language models}.
\newblock In \emph{Advances in Neural Information Processing Systems}, volume~35, pages 24824--24837. Curran Associates, Inc.

\bibitem[{Wen et~al.(2024)Wen, Li, Liu, Zhao, Yu, Li, Jiang, Liu, Zhang, and Liu}]{wen2024autodroid}
Hao Wen, Yuanchun Li, Guohong Liu, Shanhui Zhao, Tao Yu, Toby Jia-Jun Li, Shiqi Jiang, Yunhao Liu, Yaqin Zhang, and Yunxin Liu. 2024.
\newblock \href {https://doi.org/10.1145/3636534.3649379} {Autodroid: Llm-powered task automation in android}.
\newblock In \emph{Proceedings of the 30th Annual International Conference on Mobile Computing and Networking}, ACM MobiCom '24, page 543–557, New York, NY, USA. Association for Computing Machinery.

\bibitem[{Wu et~al.(2024)Wu, Zhong, Xing, Lai, Liu, Wang, Chen, Zhu, Lu, Lu, Luo, Qiao, and Dai}]{wu2024visionllmv2endtoendgeneralist}
Jiannan Wu, Muyan Zhong, Sen Xing, Zeqiang Lai, Zhaoyang Liu, Wenhai Wang, Zhe Chen, Xizhou Zhu, Lewei Lu, Tong Lu, Ping Luo, Yu~Qiao, and Jifeng Dai. 2024.
\newblock \href {https://arxiv.org/abs/2406.08394} {Visionllm v2: An end-to-end generalist multimodal large language model for hundreds of vision-language tasks}.
\newblock \emph{Preprint}, arXiv:2406.08394.

\bibitem[{Wu et~al.(2023)Wu, Wang, Xu, Lu, and Yan}]{wu2023embodiedtaskplanninglarge}
Zhenyu Wu, Ziwei Wang, Xiuwei Xu, Jiwen Lu, and Haibin Yan. 2023.
\newblock \href {https://arxiv.org/abs/2307.01848} {Embodied task planning with large language models}.
\newblock \emph{Preprint}, arXiv:2307.01848.

\bibitem[{Xi et~al.(2023)Xi, Chen, Guo, He, Ding, Hong, Zhang, Wang, Jin, Zhou, Zheng, Fan, Wang, Xiong, Zhou, Wang, Jiang, Zou, Liu, Yin, Dou, Weng, Cheng, Zhang, Qin, Zheng, Qiu, Huang, and Gui}]{xi2023risepotentiallargelanguage}
Zhiheng Xi, Wenxiang Chen, Xin Guo, Wei He, Yiwen Ding, Boyang Hong, Ming Zhang, Junzhe Wang, Senjie Jin, Enyu Zhou, Rui Zheng, Xiaoran Fan, Xiao Wang, Limao Xiong, Yuhao Zhou, Weiran Wang, Changhao Jiang, Yicheng Zou, Xiangyang Liu, Zhangyue Yin, Shihan Dou, Rongxiang Weng, Wensen Cheng, Qi~Zhang, Wenjuan Qin, Yongyan Zheng, Xipeng Qiu, Xuanjing Huang, and Tao Gui. 2023.
\newblock \href {https://arxiv.org/abs/2309.07864} {The rise and potential of large language model based agents: A survey}.
\newblock \emph{Preprint}, arXiv:2309.07864.

\bibitem[{Xie et~al.(2024)Xie, Zhang, Chen, Li, Zhao, Cao, Toh, Cheng, Shin, Lei et~al.}]{xie2024osworldbenchmarkingmultimodalagents}
Tianbao Xie, Danyang Zhang, Jixuan Chen, Xiaochuan Li, Siheng Zhao, Ruisheng Cao, Jing~Hua Toh, Zhoujun Cheng, Dongchan Shin, Fangyu Lei, et~al. 2024.
\newblock Osworld: Benchmarking multimodal agents for open-ended tasks in real computer environments.
\newblock In \emph{Advances in Neural Information Processing Systems}, volume~37, pages 52040--52094.

\bibitem[{Xing et~al.(2024)Xing, Zhang, Xue, Chen, Yang, and Xiao}]{xing2024understanding}
Mingzhe Xing, Rongkai Zhang, Hui Xue, Qi~Chen, Fan Yang, and Zhen Xiao. 2024.
\newblock Understanding the weakness of large language model agents within a complex android environment.
\newblock pages 6061--6072.

\bibitem[{Xu et~al.(2024)Xu, Chen, Zhao, Ma, Cao, Zhu, and Yu}]{10.1145/3616855.3635753}
Hongshen Xu, Lu~Chen, Zihan Zhao, Da~Ma, Ruisheng Cao, Zichen Zhu, and Kai Yu. 2024.
\newblock \href {https://doi.org/10.1145/3616855.3635753} {Hierarchical multimodal pre-training for visually rich webpage understanding}.
\newblock In \emph{Proceedings of the 17th ACM International Conference on Web Search and Data Mining}, WSDM '24, page 864–872, New York, NY, USA. Association for Computing Machinery.

\bibitem[{Xu et~al.()Xu, Zhu, Zhang, Ma, Fan, Chen, and Yu}]{xu2024rejectionimprovesreliabilitytraining}
Hongshen Xu, Zichen Zhu, Situo Zhang, Da~Ma, Shuai Fan, Lu~Chen, and Kai Yu.
\newblock \href {https://arxiv.org/abs/2403.18349} {Rejection improves reliability: Training llms to refuse unknown questions using rl from knowledge feedback}.
\newblock In \emph{First Conference on Language Modeling}.

\bibitem[{Yan et~al.(2023)Yan, Yang, Zhu, Lin, Li, Wang, Yang, Zhong, McAuley, Gao, Liu, and Wang}]{yan2023gpt4vwonderlandlargemultimodal}
An~Yan, Zhengyuan Yang, Wanrong Zhu, Kevin Lin, Linjie Li, Jianfeng Wang, Jianwei Yang, Yiwu Zhong, Julian McAuley, Jianfeng Gao, Zicheng Liu, and Lijuan Wang. 2023.
\newblock \href {https://arxiv.org/abs/2311.07562} {Gpt-4v in wonderland: Large multimodal models for zero-shot smartphone gui navigation}.
\newblock \emph{Preprint}, arXiv:2311.07562.

\bibitem[{Yao et~al.(2023)Yao, Zhao, Yu, Du, Shafran, Narasimhan, and Cao}]{yao2022react}
Shunyu Yao, Jeffrey Zhao, Dian Yu, Nan Du, Izhak Shafran, Karthik~R Narasimhan, and Yuan Cao. 2023.
\newblock \href {https://openreview.net/forum?id=WE_vluYUL-X} {React: Synergizing reasoning and acting in language models}.
\newblock In \emph{The Eleventh International Conference on Learning Representations}.

\bibitem[{Yao et~al.(2024)Yao, Yu, Zhang, Wang, Cui, Zhu, Cai, Li, Zhao, He, Chen, Zhou, Zou, Zhang, Hu, Zheng, Zhou, Cai, Han, Zeng, Li, Liu, and Sun}]{yao2024minicpmvgpt4vlevelmllm}
Yuan Yao, Tianyu Yu, Ao~Zhang, Chongyi Wang, Junbo Cui, Hongji Zhu, Tianchi Cai, Haoyu Li, Weilin Zhao, Zhihui He, Qianyu Chen, Huarong Zhou, Zhensheng Zou, Haoye Zhang, Shengding Hu, Zhi Zheng, Jie Zhou, Jie Cai, Xu~Han, Guoyang Zeng, Dahai Li, Zhiyuan Liu, and Maosong Sun. 2024.
\newblock \href {https://arxiv.org/abs/2408.01800} {Minicpm-v: A gpt-4v level mllm on your phone}.
\newblock \emph{Preprint}, arXiv:2408.01800.

\bibitem[{Ye et~al.(2024)Ye, Xu, Xu, Ye, Yan, Zhou, Wang, Hu, Shi, Shi, Li, Xu, Chen, Tian, Qian, Zhang, Huang, and Zhou}]{ye2024mplugowlmodularizationempowerslarge}
Qinghao Ye, Haiyang Xu, Guohai Xu, Jiabo Ye, Ming Yan, Yiyang Zhou, Junyang Wang, Anwen Hu, Pengcheng Shi, Yaya Shi, Chenliang Li, Yuanhong Xu, Hehong Chen, Junfeng Tian, Qi~Qian, Ji~Zhang, Fei Huang, and Jingren Zhou. 2024.
\newblock \href {https://arxiv.org/abs/2304.14178} {mplug-owl: Modularization empowers large language models with multimodality}.
\newblock \emph{Preprint}, arXiv:2304.14178.

\bibitem[{Ye et~al.(2023)Ye, Xu, Ye, Yan, Hu, Liu, Qian, Zhang, Huang, and Zhou}]{ye2023mplugowl2revolutionizingmultimodallarge}
Qinghao Ye, Haiyang Xu, Jiabo Ye, Ming Yan, Anwen Hu, Haowei Liu, Qi~Qian, Ji~Zhang, Fei Huang, and Jingren Zhou. 2023.
\newblock \href {https://arxiv.org/abs/2311.04257} {mplug-owl2: Revolutionizing multi-modal large language model with modality collaboration}.
\newblock \emph{Preprint}, arXiv:2311.04257.

\bibitem[{Zhang et~al.(2024{\natexlab{a}})Zhang, Li, He, Zhang, Qiao, Qin, Ma, Kang, Lin, Rajmohan, Zhang, and Zhang}]{zhang2024ufo}
Chaoyun Zhang, Liqun Li, Shilin He, Xu~Zhang, Bo~Qiao, Si~Qin, Minghua Ma, Yu~Kang, Qingwei Lin, Saravan Rajmohan, Dongmei Zhang, and Qi~Zhang. 2024{\natexlab{a}}.
\newblock \href {https://arxiv.org/abs/2402.07939} {Ufo: A ui-focused agent for windows os interaction}.
\newblock \emph{Preprint}, arXiv:2402.07939.

\bibitem[{Zhang et~al.(2023{\natexlab{a}})Zhang, Yang, Liu, Han, Chen, Huang, Fu, and Yu}]{yang2023appagent}
Chi Zhang, Zhao Yang, Jiaxuan Liu, Yucheng Han, Xin Chen, Zebiao Huang, Bin Fu, and Gang Yu. 2023{\natexlab{a}}.
\newblock \href {https://arxiv.org/abs/2312.13771} {Appagent: Multimodal agents as smartphone users}.
\newblock \emph{Preprint}, arXiv:2312.13771.

\bibitem[{Zhang et~al.(2023{\natexlab{b}})Zhang, Chen, Zhang, Xu, Zhao, and Yu}]{DanyangZhang2023NeurIPS_Rememberer}
Danyang Zhang, Lu~Chen, Situo Zhang, Hongshen Xu, Zihan Zhao, and Kai Yu. 2023{\natexlab{b}}.
\newblock \href {http://papers.nips.cc/paper\_files/paper/2023/hash/f6b22ac37beb5da61efd4882082c9ecd-Abstract-Conference.html} {Large language models are semi-parametric reinforcement learning agents}.
\newblock In \emph{Advances in Neural Information Processing Systems 36: Annual Conference on Neural Information Processing Systems 2023, NeurIPS 2023, New Orleans, LA, USA, December 10 - 16, 2023}.

\bibitem[{Zhang et~al.(2024{\natexlab{b}})Zhang, Shen, Xie, Zhang, Xie, Zhao, Chen, Chen, Xu, Cao, and Yu}]{zhang2024mobileenvbuildingqualifiedevaluation}
Danyang Zhang, Zhennan Shen, Rui Xie, Situo Zhang, Tianbao Xie, Zihan Zhao, Siyuan Chen, Lu~Chen, Hongshen Xu, Ruisheng Cao, and Kai Yu. 2024{\natexlab{b}}.
\newblock \href {https://arxiv.org/abs/2305.08144} {Mobile-env: Building qualified evaluation benchmarks for llm-gui interaction}.
\newblock \emph{Preprint}, arXiv:2305.08144.

\bibitem[{Zhang et~al.(2024{\natexlab{c}})Zhang, Zhang, Liu, Song, Wang, Krishna, and Wu}]{zhang2024training}
Shaokun Zhang, Jieyu Zhang, Jiale Liu, Linxin Song, Chi Wang, Ranjay Krishna, and Qingyun Wu. 2024{\natexlab{c}}.
\newblock Training language model agents without modifying language models.
\newblock \emph{ICML'24}.

\bibitem[{Zhang et~al.(2024{\natexlab{d}})Zhang, Bo, Ma, Li, Chen, Dai, Zhu, Dong, and Wen}]{zhang2024surveymemorymechanismlarge}
Zeyu Zhang, Xiaohe Bo, Chen Ma, Rui Li, Xu~Chen, Quanyu Dai, Jieming Zhu, Zhenhua Dong, and Ji-Rong Wen. 2024{\natexlab{d}}.
\newblock \href {https://arxiv.org/abs/2404.13501} {A survey on the memory mechanism of large language model based agents}.
\newblock \emph{Preprint}, arXiv:2404.13501.

\bibitem[{Zhang and Zhang(2024)}]{zhan2023autoui}
Zhuosheng Zhang and Aston Zhang. 2024.
\newblock You only look at screens: Multimodal chain-of-action agents.
\newblock In \emph{Findings of the Association for Computational Linguistics ACL 2024}, pages 3132--3149.

\bibitem[{Zhang et~al.(2023{\natexlab{c}})Zhang, Zhang, Li, and Smola}]{zhang2023automatic}
Zhuosheng Zhang, Aston Zhang, Mu~Li, and Alex Smola. 2023{\natexlab{c}}.
\newblock \href {https://openreview.net/forum?id=5NTt8GFjUHkr} {Automatic chain of thought prompting in large language models}.
\newblock In \emph{The Eleventh International Conference on Learning Representations}.

\bibitem[{Zheng et~al.(2024)Zheng, Gou, Kil, Sun, and Su}]{BoyuanZheng2024ICML_SeeAct}
Boyuan Zheng, Boyu Gou, Jihyung Kil, Huan Sun, and Yu~Su. 2024.
\newblock \href {https://openreview.net/forum?id=piecKJ2DlB} {Gpt-4v(ision) is a generalist web agent, if grounded}.
\newblock In \emph{Forty-first International Conference on Machine Learning, {ICML} 2024, Vienna, Austria, July 21-27, 2024}. OpenReview.net.

\bibitem[{Zhu et~al.(2024)Zhu, Chen, Shen, Li, and Elhoseiny}]{zhu2024minigpt}
Deyao Zhu, Jun Chen, Xiaoqian Shen, Xiang Li, and Mohamed Elhoseiny. 2024.
\newblock \href {https://openreview.net/forum?id=1tZbq88f27} {Mini{GPT}-4: Enhancing vision-language understanding with advanced large language models}.
\newblock In \emph{The Twelfth International Conference on Learning Representations}.

\bibitem[{Zhu et~al.(2023)Zhu, Sun, Yang, Peng, Zou, Li, Li, Chen, Ma, Zhang et~al.}]{zhu2023cam}
Zichen Zhu, Liangtai Sun, Jingkai Yang, Yifan Peng, Weilin Zou, Ziyuan Li, Wutao Li, Lu~Chen, Yingzi Ma, Danyang Zhang, et~al. 2023.
\newblock Cam-gui: A conversational assistant on mobile gui.
\newblock In \emph{National Conference on Man-Machine Speech Communication}, pages 302--315. Springer.

\end{thebibliography}

\newpage
\appendix


\section{Several Useful Links}
\noindent Code of \agent{}: 

\url{https://github.com/OpenDFM/MobA}

\noindent Prompts used in \agent{}:

\url{https://github.com/OpenDFM/MobA/blob/main/moba/prompts/prompts.py}

\noindent Complete \bench{}: 

\url{https://huggingface.co/datasets/OpenDFM/MobA-MobBench}


\section{View hierarchy processing}
\begin{algorithm*}
    \small
    \rule{0.9\linewidth}{1pt}

    \KwIn{xml file of the current screen}
    \KwOut{the annotated screen}
    
    \tcp{First pass: Filter the small elements and all useless attributes}
    
    elements $\gets$ (sort(filter(elements), key=area)
    
    selected\_elements $\gets \emptyset $

    \tcp{Second pass: select elements whose overlapping area with former ones is small}
    \ForEach{element in elements}{
        \If{element is interactive}{
            is\_valid $\gets$ True
            \ForEach{selected\_element in selected\_elements}{
                \If{overlapping\_area is large}{
                    is\_valid $\gets$ False
                }
            }
            \If{is\_valid is True}{
                selected\_elements $\gets$ selected\_elements + element
            }
        }
    }

    \tcp{Third pass: Add the texts and merge the information of text into interactive elements}
    \ForEach{element in elements}{
        \ForEach{selected\_element in selected\_elements}{
            \If{element is contained in selected\_element}{
                merge(element, selected\_element)
            }
        }
    }
    
    \tcp{Final pass: Sort the elements from left to right, top to bottom}
    
    Sort(elements, key=(y,x))
    
    Plot all the interactive elements with their index

    \rule{0.9\linewidth}{1pt}
    \vspace{0.5cm}
    \caption{The Logic of View-Hierarchy Process Algorithm}
    \label{alg:vh}
    
\end{algorithm*}
\ULforem

\begin{figure*}[ht]
    \centering
       \includegraphics[width=0.9\textwidth]{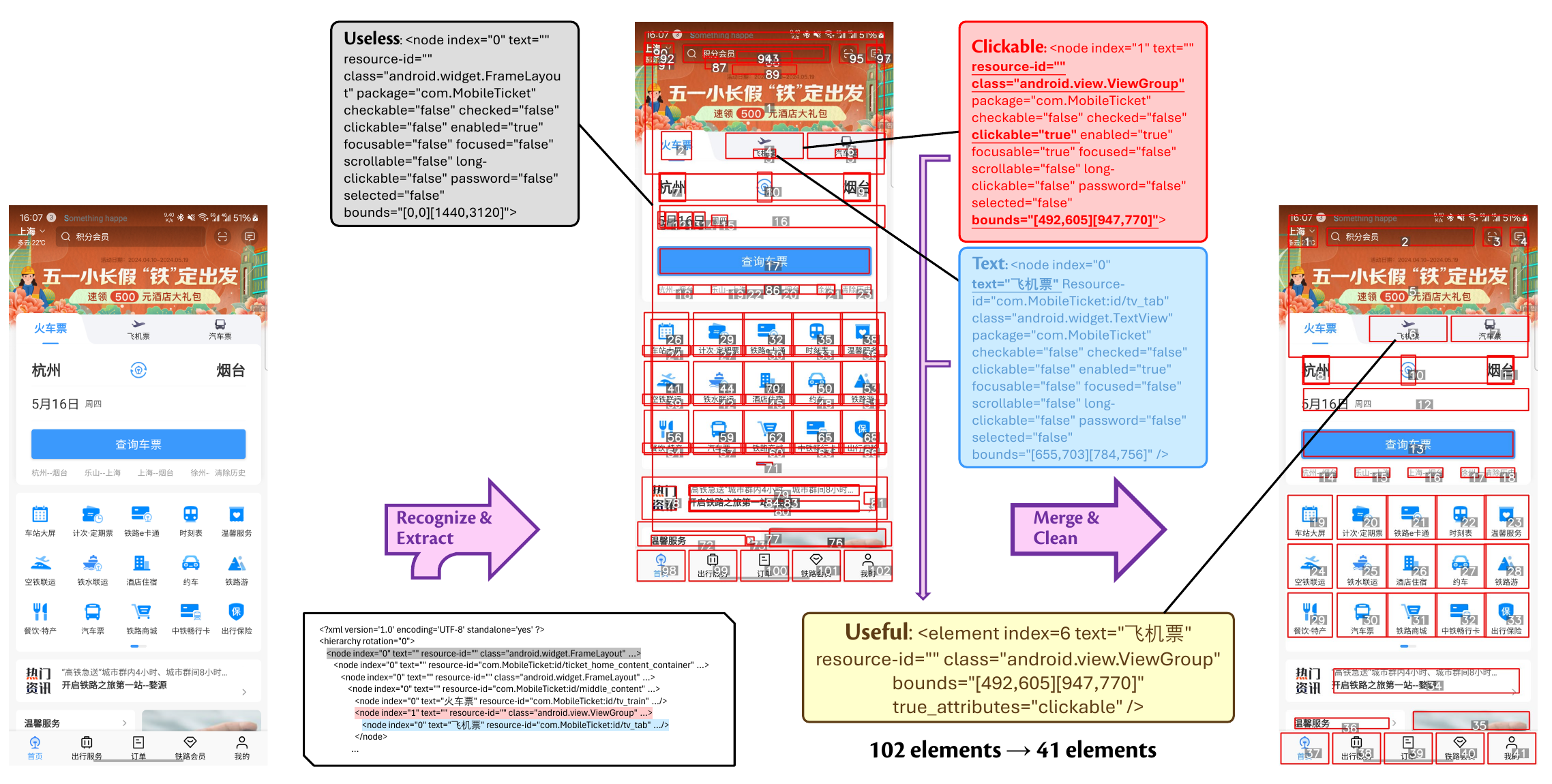}
       \caption{\textbf{An Example Diagram of View-Hierarchy Processing.} From left to right are the original image, unprocessed image and processed image. The underlined parts are the properties that are retained after the merge.}
       \label{fig:vh_proc}
\end{figure*}

Given that (1) large models still exhibit limitations in processing visual information and (2) certain elements of the mobile phone interface cannot be obtained through visual means alone, the view hierarchy (VH) plays a crucial role in enabling agents to effectively interpret the mobile interface. However, the XML files representing mobile interfaces contain a substantial amount of redundant information. This redundancy increases token counts and complicates the agent’s task of identifying key UI elements.

To address this issue, we developed an algorithm designed to filter UI elements. The algorithm consists of four steps: (1) parsing UI elements from the XML file, (2) filtering user-interactable UI elements based on their attributes, and adding them in ascending order of size, unless they exhibit significant overlap with previously added elements, (3) for UI elements containing text, merging the text content with interactive elements if the text is largely contained within those elements, thus enriching the interactive element with explanatory information, and (4) assigning an index to each UI element according to its central coordinates, from left to right and top to bottom, while plain text elements are assigned an index of -1. This ensures that the index ordering aligns more closely with the user's natural visual scanning behavior.

In summary, the core of our algorithm is the preservation of key interactive elements and their associated textual information, while minimizing occlusion in the image. For example, in the case of the "plane ticket" element demonstrated in \Cref{fig:vh_proc}, the UI element itself does not contain text, and the text information associated with the plane ticket is non-clickable. By merging the two, the agent can infer that clicking the UI element corresponds to selecting the plane ticket.

However, limitations remain in this approach. There are cases where all elements in the XML file are marked as "clickable=false", despite the presence of interactive elements in practice. Additionally, technical limitations sometimes prevent the XML file from accurately reflecting the current state of the interface.


\section{Action Space}

We provide all actions supported in \agent{} in \Cref{tab:actions}.

\begin{table*}[h!]
\footnotesize
\centering
\begin{tabular}{m{2cm}<{\raggedright} m{1.5cm}<{\raggedright} m{5cm}<{\raggedright} m{5cm}<{\raggedright} }
\toprule
\textbf{Action} & \textbf{Type}  & \textbf{Usage} & \textbf{Description} \\
\midrule
Click & single  & Click(element\_index: int) & This function clicks the center of the UI element with the specified element index.  \\
\midrule
Click by Coordinate & single & Click\_by\_Coordinate(x: double, y: double) & This function simulates a click at the specified x and y coordinates on the screen. \\
\midrule
Double Click & single  & Double\_Click(element\_index: int) & This function double clicks the center of the UI element with the specified element index.  \\
\midrule
Long Press & single  & Long\_Press(element\_index: int) & This function long-presses the center of the UI element with the specified element index. \\
\midrule
Scroll & single & Scroll(element\_index: int, direction: str, distance: str or int) & This function swipes from the center of the UI element with the specified element index.  \\
\midrule
Swipe & single  & Swipe(direction: str, distance: str) & This function swipes from the center of the screen.  \\
\midrule
Type & single & Type(text: str) & This function inputs text on the current input box.  \\
\midrule
Back & single  & Back() & This function presses the back key to return to the previous screen or status.  \\
\midrule
Box Input & combination & Box\_Input(element\_index: int, text: str) & This function clicks the input box, inputs given text, and confirms it.  \\
\midrule
Open App & system & Open\_App(description: Optional[str]) & This function locates and opens an app with a short description.  \\
\midrule
Close App & system & Close\_App(package\_name: Optional[str]) & This function closes the specified app by its package name. \\
\midrule
Error & system & Failed() & This function indicates that the task cannot be completed.  \\
\midrule
Finish & system & Finish() & This function indicates that the task is completed.  \\
\bottomrule
\end{tabular}
\caption{\textbf{Available Actions and Descriptions}}
\label{tab:actions}
\end{table*}

\section{\bench}
\label{app:bench}

We provide five examples of the tasks included in \bench~as shown in \Cref{tab:bench}. You can get the complete collection of 50 tasks in both Chinese and English on \href{https://huggingface.co/datasets/OpenDFM/MobA-MobBench}{Huggingface}.

   \begin{table*}[htbp]
        \centering
        \small
        \setlength\tabcolsep{3pt}
        \renewcommand{\arraystretch}{1.0}
        \begin{tabular}{m{1.3cm}<{\centering} m{1.8cm}<{\centering} m{4.5cm}<{\raggedright} m{2.5cm}<{\raggedright} m{3cm}<{\raggedright} m{1cm}<{\raggedright}}
        
            \toprule
            \textbf{Type}      & \textbf{Application}   & \textbf{Task}     & \textbf{Preparation}  & \textbf{Scoring Milestones} & \textbf{Steps} \\
            \midrule
            Easy  & McDonald's & Switch the language of the McDonald's app to English. & Switch to Chinese.   & 1. Task completion.  & 6.7 \\
            \midrule
            Medium  & 12306 \newline(China Railway) & Check the schedule for train G104 from Shanghai to Beijing tomorrow, and find out what time it is expected to arrive in Nanjing. & -   & 1. Enter the timetable screen, \newline2. Correct train number, \newline3. Task completion. &11.7  \\
            \midrule
            Hard & Douban & Search for the movie "The Shawshank Redemption" on Douban, mark it as "watched", rate it five stars, and leave a positive review. & Remove the previous mark, rating, and review of this movie. & 1. Correct movie, \newline2. Correct mark, \newline3. Correct rating, \newline4. Positive review. & 9.7  \\
            \midrule
            Indirect & BiliBili   & If I'm out of mobile data, what videos can I still watch on the phone?   &  Download several videos in advance.   &  1. Open BiliBili, \newline2. Check downloads. &    3.3  \\
            \midrule
            Cross-APP & JD.com, WeChat & Share the product link of the most recent JD.com order with a WeChat friend, and write a recommendation message. & There is an existing order. & 1. Enter the order list, \newline2. Correct order, \newline3. Suitable message, \newline4. Task completion. &  10.3\\
            \bottomrule
        \end{tabular}

        \caption{\textbf{Several example tasks in \bench.} The content is translated from Chinese.}
        \label{tab:bench}
    \end{table*}

\section{Detailed Results Comparison}
While the performance of all models is relatively similar on simpler tasks, \agent{} demonstrates superior results in more challenging tasks, outperforming other models except for Human and GPT-4o + Human. This suggests that \agent{} is more efficient in handling complex cases. Additionally, the incorporation of both the Memory Module and Plan Module enhances performance, highlighting their respective contributions to the system's overall capability.

\begin{figure*}[!h]
    \centering
    \includegraphics[width=1.0\linewidth,trim=0 0 0 0,clip]{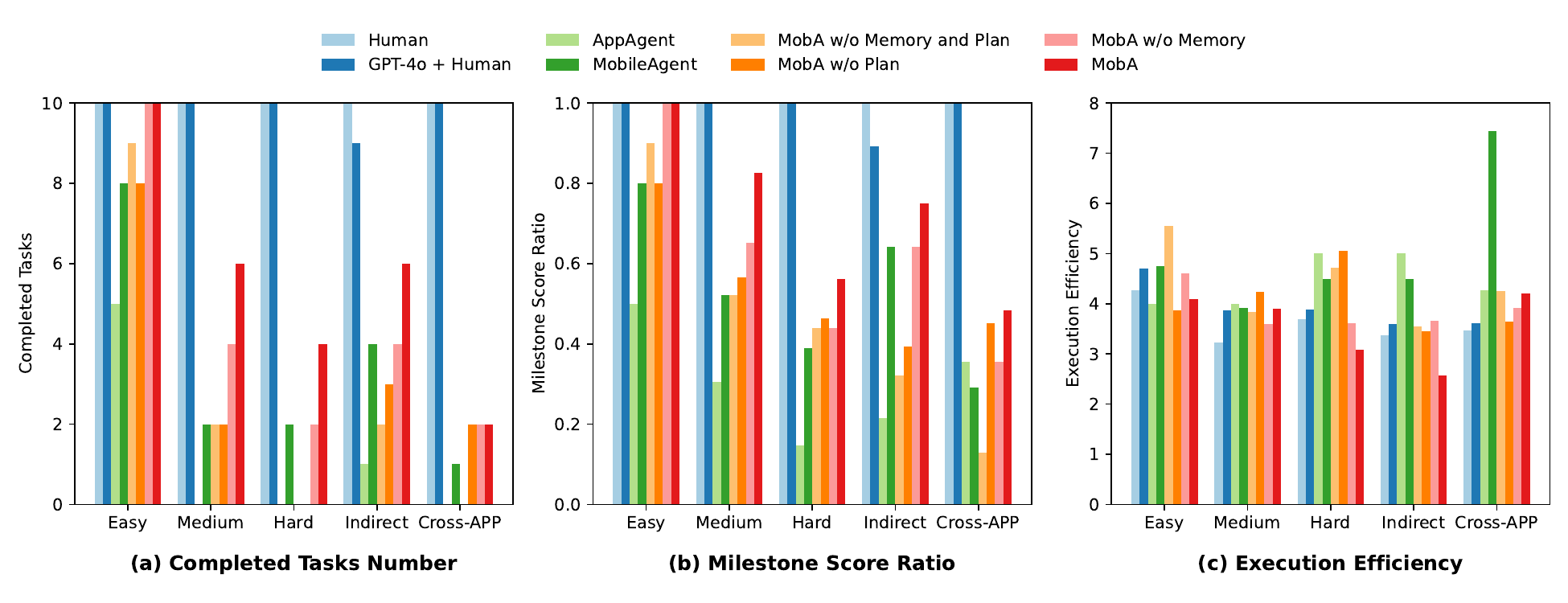}
    \caption{\textbf{Performance on \bench~Categorized by Task Type. }}
    \label{fig:result}
\end{figure*}

\subsection{Human is more adaptive and robust to screen interactions}
While the human baseline is considered the optimal solution for each task, the \textit{GPT-4o + Human} method achieves performance very close to that of human operators on all metrics. In the evaluation of \textit{GPT-4o + Human}, the agent only provides textual task descriptions and an initial screenshot, and the GPT-4o generates detailed step-by-step instructions, which are then executed manually by a human operator. 

The eye-catching performance of \textit{GPT-4o + Human} can be attributed to several factors: (1) a relatively lenient standard in task execution, allowing human operators to interpret GPT-4o's general instructions flexibly; (2) human operators automatically completing tasks such as OCR, target detection, and localization, ensuring more precise actions; (3) GPT-4o provides a global plan, avoiding redundant or missed steps; (4) technical issues (e.g., inability to retrieve XML files or missing information in the files) do not affect task completion.

\end{document}